# Atmospheric Pressure Ammonia Synthesis on AuRu Catalysts Enabled by Plasmon-Controlled Hydrogenation and Nitrogen-species Desorption

Lin Yuan[1, *], Briley B. Bourgeois[1], Elijah Begin[3], Yirui Zhang[1], Alan X. Dai[2], Zhihua Cheng[11], Amy S. McKeown-Green[5], Zhichen Xue[8], Yi Cui[1], Kun Xu[6, 7], Yu Wang[6, 7], Matthew R. Jones[11], Yi Cui[1,9,10], Arun Majumdar[6, 7], Junwei Lucas Bao[3, *], and Jennifer A. Dionne[1, 4, *]

[1]Department of Materials Science and Engineering, Stanford University School of Engineering, Stanford, CA, 94305, United States
[2]Department of Chemical Engineering, Stanford University School of Engineering, Stanford, CA, 94305, United States
[3]Department of Chemistry, Boston College, Chestnut Hill, MA, 02467, United States
[4]Department of Radiology, Stanford University School of Medicine, Stanford, CA, 94305, United States
[5]Department of Chemistry, Stanford University School of Humanities and Sciences, Stanford, CA, 94305, United States
[6]Stanford Doerr School of Sustainability, Stanford University, Stanford, CA, 94305, United States
[7]Department of Mechanical Engineering, Stanford University School of Engineering, Stanford, CA, 94305, United States
[8]Stanford Synchrotron Radiation Lightsource, SLAC National Accelerator Laboratory, Menlo Park, CA, 94025, United States
[9]Stanford Institute for Materials and Energy Sciences, SLAC National Accelerator Laboratory, Menlo Park, CA, 94025, United States
[10]Department of Energy Science and Engineering, Stanford University, Stanford, CA, 94305, United States
[11]Department of Chemistry, Rice University, Houston, TX, 77025, United States

* Corresponding authors:
ly31@stanford.edu;
lucas.bao@bc.edu;
jdionne@stanford.edu.

## Abstract

Ammonia is a key component of fertilizer and can potentially be directly utilized as a clean fuel and a hydrogen carrier. The Haber-Bosch process for ammonia synthesis consumes more than half of the annual industrial hydrogen and contributes up to ~3% of global greenhouse gas emissions. Light-driven chemical reactions assisted by surface plasmon resonances could provide an alternative, less energy-intensive pathway for the production of ammonia by altering the reaction intermediates via non-equilibrium processes. Here, we report gold-ruthenium plasmonic bimetallic alloys for ammonia synthesis at room temperature and pressure, driven solely by visible light illumination. We use colloidal synthesis to create AuRu$_x$ alloys (x=0.1, 0.2, 0.3) and disperse these nanoparticles on MgO supports for gas-phase ammonia synthesis. We observe a ~60 µmol/g/h reactivity and ~0.12% external quantum efficiency on a AuRu$_{0.2}$ sample under 100 mW/cm$^2$ visible light illumination. *In-situ* diffuse reflective infrared Fourier transform spectroscopic measurements show the kinetics of hydrogenation of nitrogen adsorbates is accelerated under light illumination compared to thermocatalysis. By combining our wavelength-dependent reactivity and spectroscopic findings with semi-classical electromagnetic modeling, we show that plasmonic bimetallic alloys can facilitate ammonia synthesis by expediting the hydrogenation of adsorbed nitrogen species via plasmon-mediated hot electrons. Fully quantum mechanical calculations reveal that the hydrogen-assisted splitting of N$_2$ in the excited state is the key mechanism for the reaction activated at ambient conditions. Therefore, light alone or H$_2$ alone cannot achieve N$_2$ dissociation – the key bottleneck to breaking the triple bond of N$_2$. Our findings are consistent with recent hypotheses of how nitrogenase enzymes catalyze ammonia production at mild conditions and provide the mechanistic understanding foundational for sustainable photochemical transformations.

## Introduction

In the early 20th century, the discovery of the Haber-Bosch process revolutionized the production of synthetic ammonia and has since made a tremendous impact on humanity. It is estimated that half the world's population is supported by Haber-Bosch enabled agriculture,[1] and nearly 50% of the nitrogen found in humans (mainly via amino acids) can be traced back to the Haber-Bosch process. There is also growing interest in ammonia as a potential hydrogen carrier and carbon-free fuel, as it is easier to liquify and transport than hydrogen gas. [2-4] Yet, the incredible societal benefit of synthetic ammonia comes with a detrimental environmental cost. Operated at temperatures of 300-500 °C and at high pressures (100-200 atm), the Haber-Bosch process also consumes 50-60% of all industrial hydrogen. These factors collectively account for ~2% of annual global energy use, and emission of ~3% of global greenhouse gases[5,6] Both current and future uses of ammonia are in critical need of less energy-demanding synthetic pathways for more sustainable foods and fuels.

Plasmonic catalysis provides a sustainable approach for ammonia synthesis in mild conditions. Localized surface plasmon resonance (LSPR) at the surface of small nanoparticles (such as Au, Ag, Cu, and Al) concentrate visible light energy, and enable chemical transformations at catalytic interfaces including Ru, Fe, and oxide, using nitrogen and hydrogen/water as the reagents.[7-11] Light energy can be converted into chemical energy through LSPR decay via generation of non-equilibrium hot carriers, strong electric field, and the photothermal effect.[12-17] Ammonia production in an ambient environment (near room temperature and pressure) has recently been demonstrated with plasmonic catalysts, achieving production rates up to ~300 µmol/g$_{cat}$/h.[18,19] However, these pioneering reports did not clarify the underlying reaction mechanisms en-route to sustainable ammonia synthesis. Important mechanistic questions include how hot carriers can get harvested into the chemical interface, and which elementary steps can be tailored by hot carriers. Addressing these questions will help guide the design of efficient catalyst nanostructures for sustainable ammonia synthesis.

In this work, we design a bimetallic alloy plasmonic catalyst, using Au as a light harvester and Ru as the catalytic component (Fig. 1a).[20] We investigate the impact of alloy composition on photochemistry by synthesizing three molar ratios of Au to Ru from 1:0.1 to 1:0.3. We find that the 0.2 molar ratio exhibits the highest quantum yield, as it optimizes both optical absorption and catalytic site density. We also compare the reaction rates of photocatalysis with thermocatalysis, showing that non-thermal (eg, hot-carrier-driven) effects enhance photo-reactivity at lower temperatures than the heat-driven case. *In-situ* diffuse reflective infrared Fourier transform spectroscopic (DRIFTS) measurements suggest the hydrogenation of nitrogen species is accelerated under light illumination; notably, $NH_3^*$ adsorbates densities show a linear increase with illumination power, whereas heating alone results in a non-monotonic change in the adsorbate density. Finally, we simulate the reaction pathway along both the ground-state and excited-state on direct dissociation and hydrogen-assisted dissociation of $N_2$ via the embedded correlated wavefunction (ECW) theory. Here, we find that the rate-determining step of the hydrogen-assisted dissociation pathway could be overcome with a ~1 eV activation energy that is accessible from visible plasmons. Taken together, our work shows that plasmons improve the kinetics of ammonia synthesis via hot-electron-controlled hydrogenation and desorption of intermediate nitrogen species.

## Results and Discussion
### Characterization

We synthesize colloidal, polycrystalline AuRu alloys with molar ratios of 1:0.1, 1:0.2, and 1:0.3 (hereafter $AuRu_{0.1}$, $AuRu_{0.2}$, and $AuRu_{0.3}$) following a previously published protocol with slight modifications.[21,22] We prepare different composition alloys to isolate the role of optical absorptivity and catalytic activity. The concentration of the plasmonically active Au component gives rise to increased optical absorption. The concentration and the distribution of the catalytic Ru component can impact the activity of the final catalyst as specific Ru structures ($B_5$ clusters) are predicted to promote nitrogen hydrogenation over others.[23] Finally, Au and Ru alloys can be found in two different polymorphs - a hexagonal close-paced (HCP) and a face-centered cubic (FCC) phase. Care must be taken to produce comparable structures when varying the alloy composition. We choose to investigate the FCC version due to stronger plasmon intensities and narrower bandwidth.[24] Our synthesis yields an average size distribution centered around 15-20 nm (a detailed size distribution is provided in Fig. S1 and representative TEM images in Fig. S2).

Figure 1b-d show representative atomic-resolution high-angle annular dark-field scanning transmission electron microscopic (HAADF-STEM) images of $AuRu_{0.1}$ (Fig. 1b and 1c), $AuRu_{0.2}$ (Fig. 1d), and $AuRu_{0.3}$ (Fig. 1e). Spatial energy dispersive spectroscopy maps, created with scanning transmission electron microscopy (STEM-EDS), confirm spatially-homogeneous alloying of three AuRu alloys (Fig. 1f and Fig. S3). The high-resolution transmission electron microscopic (HRTEM, Fig. S4) images of $AuRu_{0.1}$, $AuRu_{0.2}$, and $AuRu_{0.3}$ show polycrystalline domain formation. The corresponding fast Fourier transform (FFT) and selected area electron diffraction (SAED) data (Fig. S4) also confirm the {111}, {200}, and {220} facets. The atomic ratio of Au and Ru in bulk is also confirmed by Inductively Coupled Plasma Optical Emission Spectroscopy (ICP-OES) measurements (results in Table S1). The X-Ray Diffraction (XRD) patterns of the as-synthesized nanocrystals with different Ru ratios show similar patterns of {111}, {200}, {220}, and {311} to the pure Au nanocrystals (Fig. S5). However, the broadening of the peaks and slight shift of 2θ angles suggest that Ru intercalates the Au FCC lattice.

Normalized extinction spectra of $AuRu_{0.1}$, $AuRu_{0.2}$, and $AuRu_{0.3}$ all show dipole plasmon peak centered near 550 nm (Fig. 1g). With increasing Ru concentration, the extinction spectra blue shifts while the peak intensity relative to the background decreases, due to the optical losses induced by Ru. We use finite difference time domain (FDTD) simulations (via Ansys Lumerical FDTD) to simulate the extinction cross-section of the different alloys (Fig. 1h). A representative near electric field enhancement under unpolarized irradiation is included in the inset of Fig. 1h. The resulting simulations show good agreement (Fig. 1g vs 1h) with the experimental trends observed in the UV-Vis spectroscopy measurements. Finally, Fig. 1i shows the possible surface mechanisms of ammonia synthesis. Multiple bond activation and formation steps can be potentially enhanced by plasmon chemistry through multiple interfacial energy/charge transfers, including electronic excitations.[3,25,26] These include 5 possible steps: 1) Nitrogen molecule splitting ($N_2+2^* \rightarrow 2N^*$); 2) Hydrogen molecule dissociation on the surface ($H_2+2^* \rightarrow 2H^*$); 3) First N-H bond formation on the surface ($N^*+H^* \rightarrow NH^*+^*$); 4) Second N-H bond formation ($NH^*+H^* \rightarrow NH_2^*+^*$); 5) Complete nitrogen hydrogenation on the surface ($NH_2^*+H^* \rightarrow NH_3^*+^*$); and 6) Ammonia desorption ($NH_3^* \rightarrow NH_3+^*$). It is these steps we wish to probe in our study. $X^*$ represents the molecule adsorbed on the metallic surface in forms of transition state, and * represents the surface sites for the molecules.

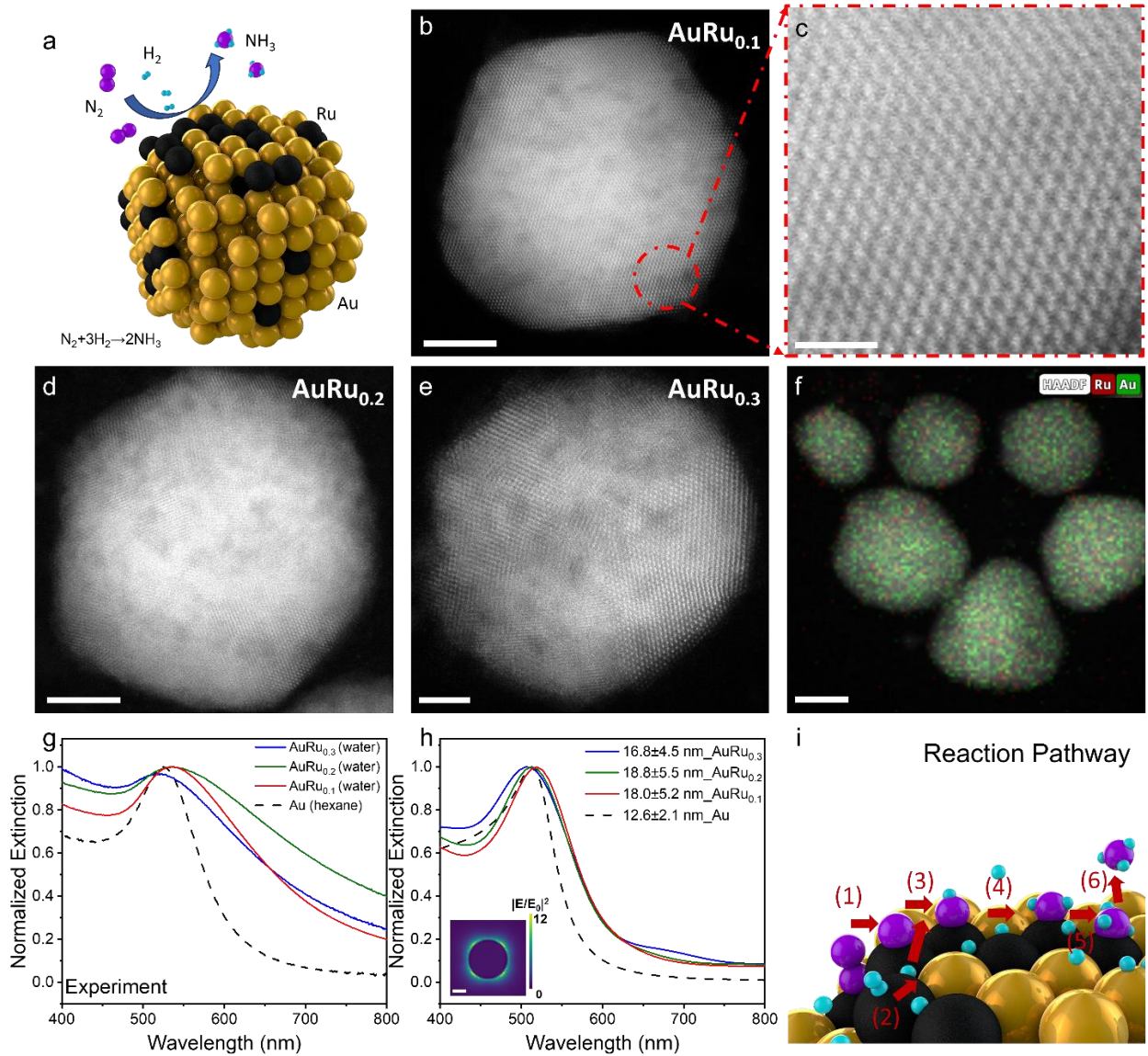

**Figure 1| Schematic of light-driven ammonia synthesis and its surface mechanism, TEM, and optical properties.** (a) A schematic diagram of the ammonia synthesis over an AuRu bimetallic alloy. (b) Atomic resolution high-angle annular dark-field scanning transmission electron microscopic (HADDF-STEM) images of $AuRu_{0.1}$ nanocrystal. (c) A zoom-in of the region of the $AuRu_{0.1}$ in (b). (d and e) HADDF-STEM images of $AuRu_{0.2}$ and $AuRu_{0.3}$ images. (f) An energy dispersive x-ray spectroscopy (STEM-EDS) mapping of Au and Ru elements on $AuRu_{0.2}$ nanoparticles. (g-h) Normalized extinction spectra for (g) experiment and (h) finite-difference time-domain (FDTD) based on effective medium approximation of the alloying. Inset of (h): An electric near-field plot of $AuRu_{0.2}$ along the plane perpendicular to the incident light at 510 nm from an FDTD simulation. (i) Schematic of surface intermediate steps for bond activation processes of ammonia synthesis. Steps (1-6): 1. Nitrogen dissociation; 2. Hydrogen dissociation; 3. Initial nitrogen hydrogenation; 4. Second N-H bond formation; 5. Complete hydrogenation; 6. Ammonia desorption. Scale bars: 1 nm for (c), and 5 nm for the rest.

## Photocatalytic measurements of supported ensemble AuRu nanoparticles

To make a heterogeneous catalyst bed, the colloidal AuRu$_{0.1}$, AuRu$_{0.2}$, and AuRu$_{0.3}$ were supported on commercial MgO powders by wet impregnation methods (Methods), and further annealed in a furnace in air to burn off the ligands. The 15 mg 3 wt% AuRu/MgO were loaded into a commercial stainless steel Harrick cell for catalytic measurements (schematic of the setup in Fig. S6). 5 sccm N$_2$ and 15 sccm H$_2$ (Praxair Inc.) were constantly flowed into the reaction chamber. The catalyst bed was heated up to 350 °C and maintained 1 hr in the same gas environment as reaction conditions, to remove the possible oxide surface states on AuRu. The downstream of the gas line was connected with an acid trap (10 mL 0.1 M HCl aqueous solution) to collect the ammonia product. The reaction chamber was either heated or irradiated by a white supercontinuum laser equipped with a bandpass filter (SuperK Varia tunable filter) to select different wavelengths (NKT Phtonics Inc.) at 100 mW/cm$^2$ (~1 solar power density) to drive the ammonia synthesis reaction (N$_2$+3H$_2$→2NH$_3$). An infrared camera (FLIR A700) above the reaction chamber recorded the real-time surface temperature of the catalyst bed through a ZnSe window. The ammonia product was detected and quantified by the Nessler method and we further confirmed the accuracy of the ammonia production by both the indophenol blue method and $^1$H NMR (Fig. S7 and S8).[27-29] More details of the measurements can be found in the Methods section.

As seen in Fig. 2a, the thermocatalytic reactivity of AuRu$_{0.1}$, AuRu$_{0.2}$, and AuRu$_{0.3}$ increases with temperature from 150-350 °C and follows the trend of the upper branch of thermal equilibrium on Ru catalysts. The reactivity also increases with the molar ratio of Ru. The wavelength-dependent reactivities of AuRu alloys from 490-700 nm illumination shows a slight, but significant peak of the reactivity at 490-550 nm, overlapping with the plasmon resonance (Fig. 2b). In contrast to the thermal-driven case, the light-driven reactivity does not scale monotonically with Ru content. Instead, the reactivity follows AuRu$_{0.2}$>AuRu$_{0.1}$>AuRu$_{0.3}$. In high-Au alloys (AuRu$_{0.1}$ & AuRu$_{0.2}$), the reactivity reaches around twice that of the thermal reaction rates despite a lower recorded surface temperature (Fig. 2c). In the highest Ru alloy (AuRu$_{0.3}$), the reactivity is lower for the light-driven reaction than the heat-driven reaction, though the light-induced heating is less than the pure thermal case. These observations are consistent with recent reports on a partial hydrogenation of acetylene reaction(C$_2$H$_2$+H$_2$→C$_2$H$_4$) on AgPd bimetallic alloy system.[30,31] Therefore, there appears to be a 'goldilocks' effect originating from the balance of light harvesting and catalytic abilities of Au and Ru, respectively. Fig. 2c shows the IR camera measured surface temperature for both light-driven and heat-driven reactions. All photoreactions show appreciable or enhanced reactivity at lower temperatures than the heat-driven case, suggesting that non-thermal, plasmon-mediated hot carriers are contributing to the ammonia synthesis reaction.[3,32]

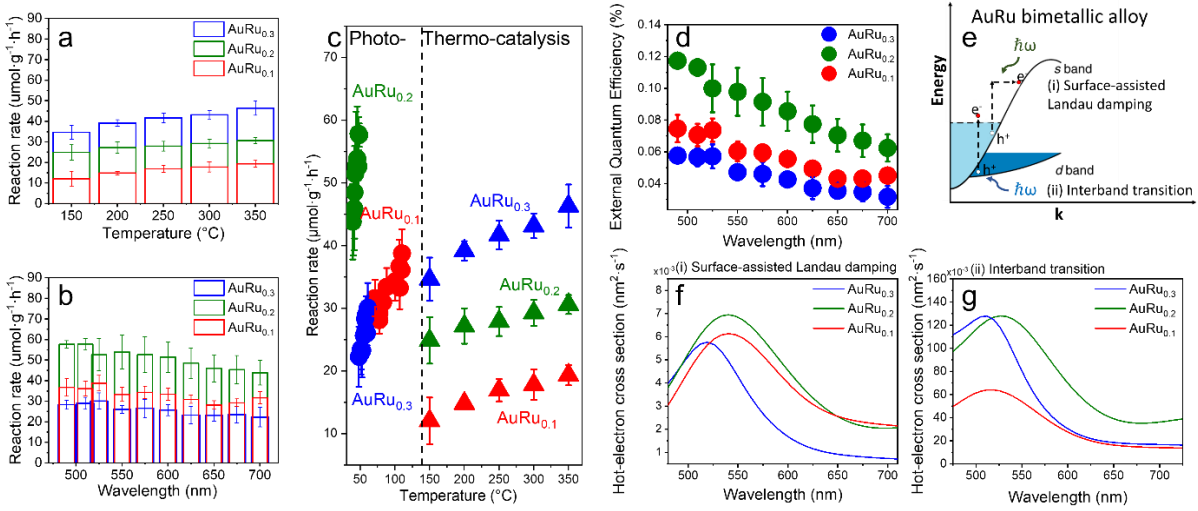

**Figure 2| Photocatalytic reactivities, and hot carrier cross sections.** (a) Thermocatalytic ammonia production rates. (b) Photocatalytic ammonia production rates under different wavelengths at 100 mW/cm$^2$. (c) Comparison of the reactivity from both photo- and thermos- catalysis with surface temperature monitored by an infrared camera. (d) Wavelength-dependent external quantum efficiency during chemical reactions. (e) Schematic of the plasmon-induced hot-carrier generation pathways from Landau damping and the interband transition. (f) Wavelength-dependent hot-carrier cross section from Landau damping. (g) Wavelength-dependent hot-carrier cross section from the interband transition.

Fig. 2d plots the external quantum efficiency (EQE) for each alloy, taking into account a 3-electron transfer per NH$_3$ molecule produced (calculations in SI). The best-performing photocatalyst, AuRu$_{0.2}$, achieves an EQE of ~0.12%, which is among the highest values compared to state-of-the-art plasmonic catalysts (up to 0.13% without sacrificial reagent, and 1% with sacrificial agent).[33] The process can be approximated as two distinct steps: (i). light-induced generation of hot carriers, then (ii). a certain portion of those hot carriers participate in transfer charge to surface adsorbates and participate in chemistry. We employed a previously developed semi-classic electromagnetic model to account for the above-mentioned two steps.[32,34] The frequency-dependent hot electron generation rate (R$_{\text{hot carrier}}$) can be calculated as:

$$R_{\text{hot carrier}}(\omega) = \frac{\delta\omega}{(\hbar\omega)^2} \int_V^{\text{MFP}} dV\, Q^{\text{MFP}}(\omega) \cdot \frac{\text{Im}(\varepsilon_{\text{surface/interband}}(\omega))}{\text{Im}(\varepsilon(\omega))} \quad (1)$$

Here, δω is the electron transition linewidth, and Q$^{\text{MFP}}$(ω) is the local absorption at the surface of the metal within the electron mean free path. ε$_{\text{interband}}$(ω) can be fitted with a Drude-Lorentz model, and ε$_{\text{surface}}$(ω) represents the dielectric permittivity contribution from surface-assisted Landau damping and can be reduced to a simple formula for a spherical model by simply introducing a Kreibig's term. Q$^{\text{MFP}}$(ω) and ε$_{\text{surface}}$(ω) can be written as:

$$Q^{\text{MFP}}(\omega) = \frac{\omega \cdot \text{Im}(\omega)}{2} \int_V^{\text{MFP}} dV\, |E(\vec{r})|^2 \quad (2)$$

$$\varepsilon_{\text{surface}}(\omega) = \frac{\omega_p^2}{\omega(\omega - i\gamma_{\text{Drude}})} - \frac{\omega_p^2}{\omega(\omega - i(\gamma_{\text{Drude}} + \frac{3}{2} \cdot \frac{\hbar v_f}{a}))} \quad (3)$$

Here $|E(\vec{r})|^2$ is the local electric field enhancement, $\omega_p$ is the bulk plasma frequency of the metal, $v_f$ is the Fermi velocity of the metal, and a is the diameter of the spherical nanoparticles. Using formulas (1-3), we can calculate the hot carrier cross-section of the AuRu, which corresponds to the first step of the light-driven reaction mechanism (Fig. S9). We assume that the majority of charge transfer occurs on Ru sites (i.e., Au is inactive for this reaction), and that there is an equal probability of charge transfer between alloys and different hot electrons. As such, we can simply weight the Ru ratio with the hot carrier cross-section to account for the second step, where the total hot carrier $R_{\text{reaction}}(\omega)$ can be written as:

$$R_{\text{reaction}}(\omega) = R_{\text{hot carrier}}(\omega) \cdot \text{Ru\%} \quad (4)$$

From equ. (1-4), we can calculate both $R_{\text{reaction}}(\omega)$ from Landau damping and interband transitions (pathways (i) and (ii) illustrated in Fig. 2e). Fig. 2f plots the theoretical scaling of the $R_{\text{reaction}}(\omega)$ for $AuRu_{0.1}$, $AuRu_{0.2}$, and $AuRu_{0.3}$ from surface-assisted Landau-damping. As seen, the trend matches with the scaling of the experimental EQE in that $AuRu_{0.2} > AuRu_{0.1} > AuRu_{0.3}$ at wavelengths above 550 nm, quantitatively explaining that the composition-dependent reactivity originates from the balance of plasmonic and catalytic components. We note that the predicted scaling from the interband transitions does not match experimentally observed results (Fig. 2g). Together, these results suggest that surface-assisted Landau damping is a bigger contributor to non-thermal reaction in small nanoparticles than interband transitions.[34,35] The misalignment from the peak position between experiment and theory (Fig. 2d vs. 2f) can be explained in part by the coulomb-screening effect in the bimetallic alloy.[36]

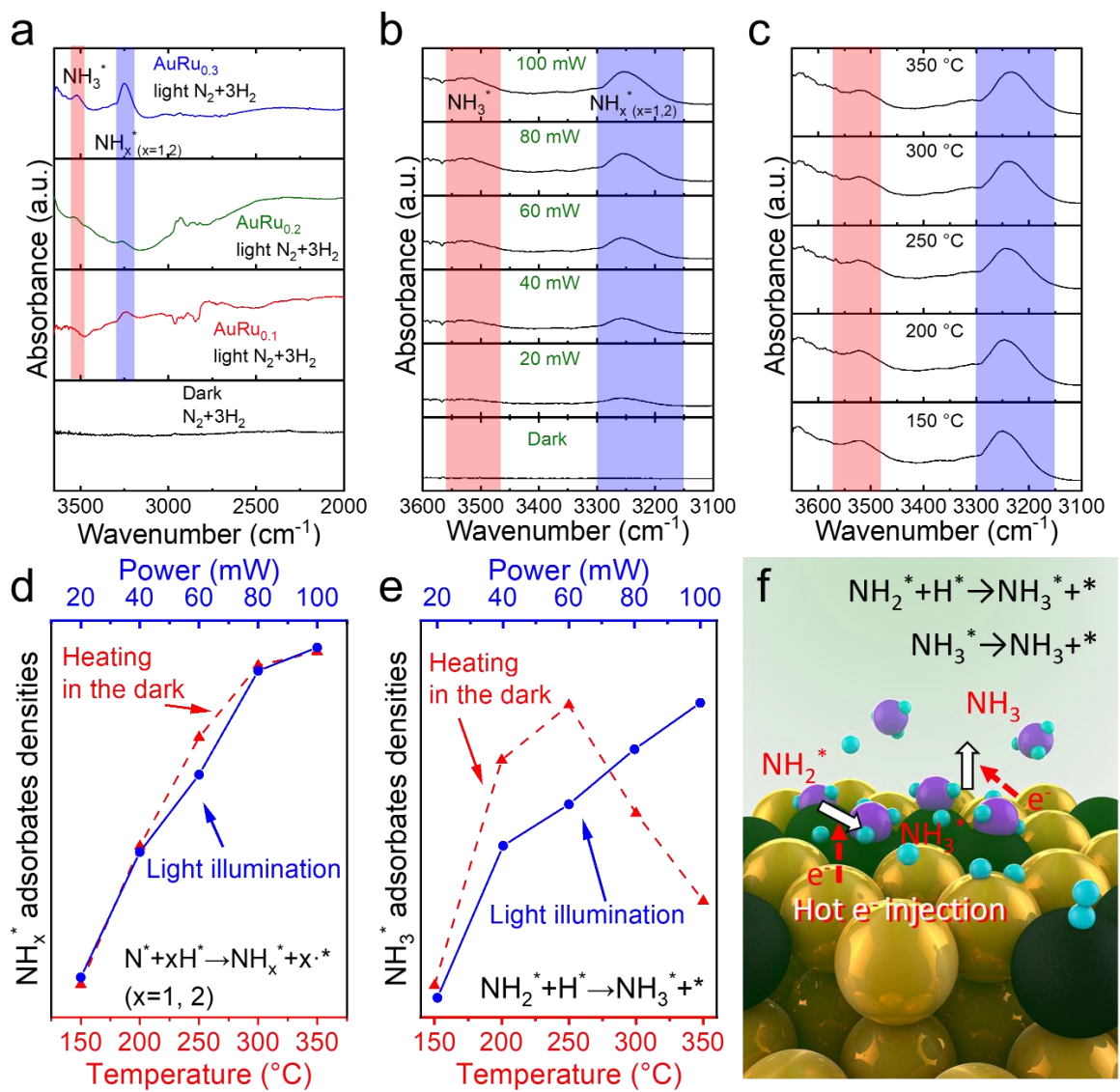

**Figure 3|** *In-situ* Diffuse Reflective Infrared Fourier Transform Spectroscopy (DRIFTS) (a) Representative *In-situ* DRIFTS spectra during ammonia synthesis under dark and illumination (525 nm, and 100 mW/cm$^2$) on AuRu$_{0.1}$, AuRu$_{0.2}$, and AuRu$_{0.3}$ nanocrystals. (b-c) Evolution of NH$_x$*(x=1,2) and NH$_3$* surface adsorbates under (b) light irradiation under 525 nm with increasing power and (c) heating in the dark on AuRu$_{0.2}$ with increasing temperature. (d) NH$_x$* and (e) NH$_3$* adsorbates densities as a function of heating temperature (thermocatalysis, red dots and lines with bottom x-axis) and power of irradiation (photocatalysis, blue dots and lines with top x-axis). The adsorbates densities are normalized so that the maximum values are the same. (f) Schematic of the proposed surface mechanism: hot-electron-induced hydrogenation and desorption accelerate the reaction.

## *In-situ* Diffuse Reflectance Infrared Fourier Transform Spectroscopy (DRIFTS) measurements and quantum mechanistic insights

We employed an *in-situ* DRIFTS setup to understand how the thermal-/light- driven intermediates evolved during ammonia production. The reactor and experimental conditions (gas flow, temperature range, illumination conditions, and catalysts bed) are the same as the previous photocatalytic or thermocatalytic measurements. Wavelength-, power-, and temperature-dependent DRIFTS spectra were recorded for each alloy system (Fig. S10 to S12). In Fig. 3, we present the representative spectra from *in-situ* DRIFTS measurements on AuRu$_{0.2}$. The AuRu$_{0.1}$ and AuRu$_{0.3}$ show similar trends (Figs. S10 to S16), which are also evidenced by the ex-situ X-ray absorption measurements where all the Ru have the similar coordination environments when doped into Au (Fig. S13). We observed two peaks centered near 3150 cm$^{-1}$ and 3550 cm$^{-1}$ in both photocatalysis and thermocatalysis for all alloys but not in the dark at room temperature on all alloys(Fig. 3a). We assigned these two peaks as NH$_x^*$(x=1, 2) and NH$_3^*$ adsorbates, respectively, according to the literature on oxide-supporting Ru catalysts.[8,37] The evolution of the two peaks under different powers (525 nm, 20-100 mW) and different temperatures range (150-350 °C) on AuRu$_{0.2}$ is highlighted in Fig. 3b and 3c, respectively. We evaluate the peak areas to draw more quantitative comparisons between the different conditions of the adsorbate concentrations in Fig. S10 to S12.

The ammonia synthesis on AuRu bimetallic alloy can be regarded as the following elementary steps[38,39] (as illustrated in Fig. 1i):

$$N_2 + 2* \rightleftharpoons 2N^* \text{ (step 1)}$$

$$H_2 + 2* \rightleftharpoons 2H^* \text{ (step 2)}$$

$$N^* + H^* \rightleftharpoons NH^* + * \text{ (step 3)}$$

$$NH^* + H^* \rightleftharpoons NH_2^* + * \text{ (step 4)}$$

$$NH_2^* + H^* \rightleftharpoons NH_3^* + * \text{ (step 5)}$$

$$NH_3^* \rightleftharpoons NH_3 + * \text{ (step 6)}$$

We used the * to represent a free surface site, and X* represents the intermediate adsorbate X on a surface site. As the peak area is proportional to the densities of surface adsorbates, the *in-situ* DRIFTS (Fig. 3) track the changes in NH$_x^*$ and NH$_3^*$ abundance during steps (4-6).

Fig. 3d and 3e compare the scaling in the relative concentrations of the intermediate species between the heat- and light-driven reactions. As seen in Fig. 3e, the NH$_x^*$ species increase with temperature, but the NH$_3^*$ species reaches a maximum at around 250 °C. This maximum could be caused by the decomposition of NH$_3^*$ at elevated temperatures, increasing the back reaction and lowering the overall reaction rate. In contrast, both the NH$_x^*$ and NH$_3^*$ species increase for the light-driven reaction, as seen in Fig. 3d, 3e, S14, and S15. It is not possible to quantitatively compare the light- and heat-driven reactions as the geometric volume of photoexcitation is certainly smaller than the catalyst bed, and possibly influences the strength of the signal measured. However, this difference in concentration trends, coupled with the observation that the light-driven reaction achieves a higher reaction rate under the same conditions, strongly suggests that there is a non-thermal contribution to the reaction, allowing the synthesis of $NH_3$ to proceed at lower temperatures. The wavelength-dependent $NH_x^*$ and $NH_3^*$ densities show a peak of 500-550 nm on all the AuRu bimetallic alloys (Fig. S16), which also indicates the plasmon-mediate hot carrier tailors the hydrogenation process of intermediate nitrogen adsorbates (Fig. 3f).

To further understand the microscopic reaction mechanism, we employed the ECW theory to calculate the reaction pathway under ground-state and excited states of ammonia synthesis, i.e., calculating the projection of the potential energy surface (PESs) along minimum energy paths (MEPs, the energy as the function of the reaction coordinate). We first simulate an optimized geometry of the {111}-facet of the Ru-doped Au surface (Fig. S17). The ex-situ X-ray Absorption Near Edge Structure (XANES) and Extended X-ray Absorption Fine Structure (EXAFS) measurements reveal that the coordination environments of Ru of all AuRu alloys are similar in this article (Fig. S13). Therefore, the simulated surface structures should represent all experiments well. The ground state calculation shows that step 3 has an activation barrier of ~0.5 eV (Fig. S18), and steps 5 and 6 have an activation barrier of ~0.2 eV and ~1.5 eV, respectively (Fig. S19). Such barriers can be overcome by the plasmon-mediated hot electrons on AuRu. Light-driven reaction potentially facilitates the forward reaction of ammonia synthesis through enhanced hydrogenation compared to thermocatalysis, resulting in higher $NH_3^*$ adsorbate densities on the surface (Fig. 3e) and reactivity (Fig. 2c).

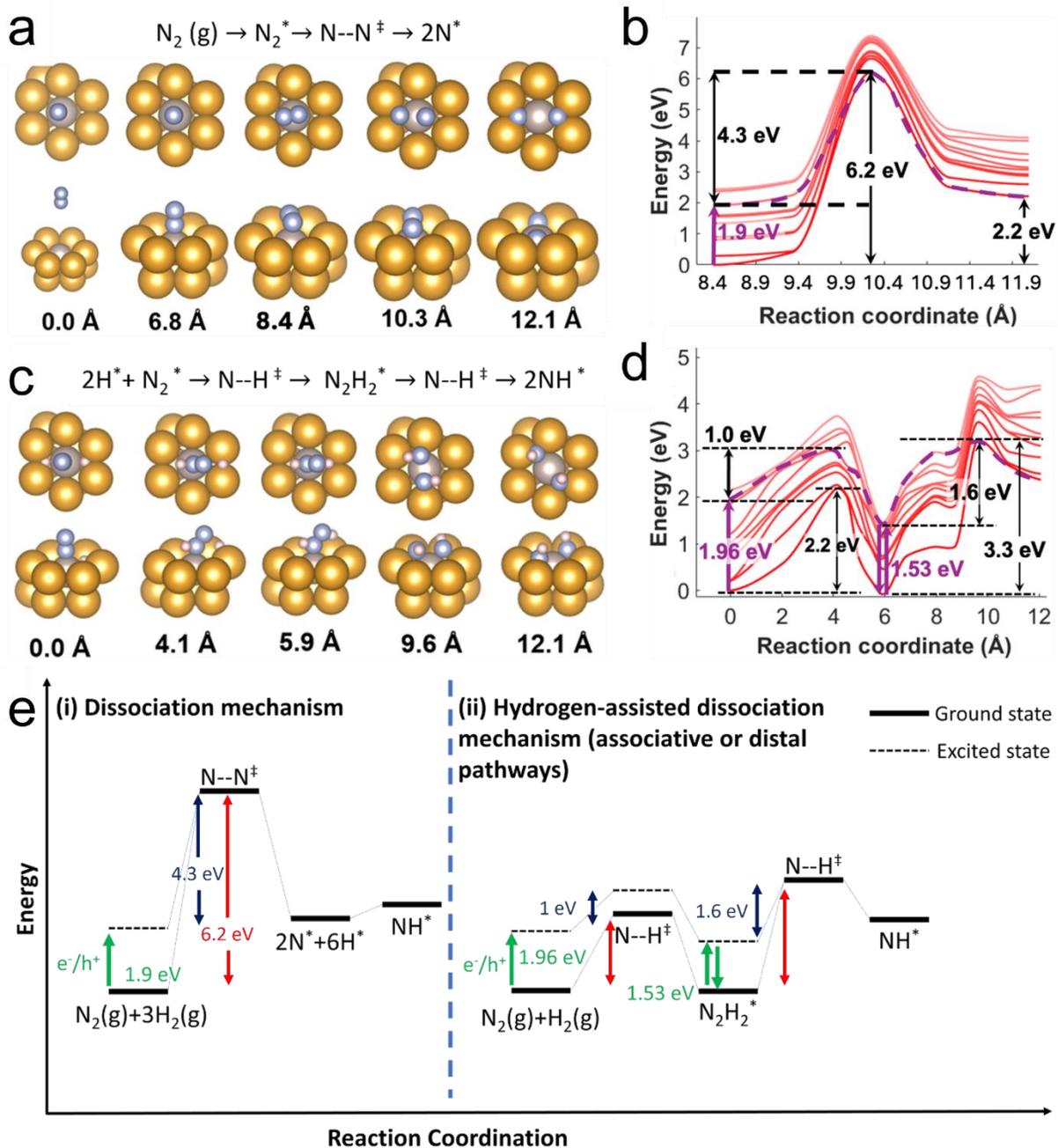

**Figure 4| Quantum mechanical simulations for reaction mechanisms.** (a) The top and side views of the optimized critical structures on the ground-state minimum-energy paths (MEPs) with the corresponding reaction coordinates for $N_2$ chemisorption. (b) The relative energies computed at the Embedded-Correlated-Wavefunction (ECW) emb-CASPT2 level for surface-bound $N_2$ dissociation $N_2$ (ads, horizontal) → 2N (ads), including both the ground state and excited states. The solid arrows represent the LSPR absorption and vertical excitation. The dashed lines represent the reactive pathways on the excited-state potential-energy surfaces. (c) The top and side views of the optimized critical structures on the ground-state MEP with the corresponding reaction coordinates for H-facilitated $N_2$ dissociation. (d) The relative energies computed at the ECW emb-CASPT2 level for surface-bound H-facilitated $N_2$ dissociation, including both the ground state and excited states. The

solid arrows represent the LSPR absorption and vertical excitation/de-excitation. The dashed lines represent the reactive pathways on the excited-state potential-energy surfaces. (e) Simplified schematic of the reaction pathways through (i) dissociation mechanism and (ii) hydrogen-assisted dissociation mechanism. Solid bars are the ground-state energy levels, while dashed bars are possible excited states accessible through plasmon-enhanced reaction pathways.

To gain a physical picture of how the rate determining steps changed under light illumination and how the hot-electron assisted hydrogenation and desorption help facilitate the bond activation and conversion, we calculated the reaction pathways, on both ground state and excited state, for the direct dissociation of $N_2$ ($N_2 \rightarrow 2N^*$), the hydrogenation of dinitrogen adsorbates ($N_2 + 2H^* \rightarrow N_2H_2^*$), and its assistance of dinitrogen splitting ($N_2H_2^* \rightarrow 2NH^*$) as the representative steps for the dissociation mechanism, associative activation, and distal pathways, respectively.[20] The simulated dissociation pathways are illustrated in Fig. 4a. The ground- and all possible excited-states potential energy diagram (Fig. 4b) suggests the activation energies of the step on the surface can be reduced from 6.2 eV to 4.3 eV, which is still inaccessible by the room temperature and pressure. The hydrogen-assisted (associative and distal mechanisms) activation of dinitrogen molecules (Fig. 4c and 4d) shows the activation is accessible through excited states, and the activation energies are reduced up to 1 eV and 1.6 eV from 2.2 eV and 3.3 eV respectively, which are accessible from visible light irradiation from the experiment. The simplified potential energy diagram from our calculations is illustrated in Fig. 4e (information from all the calculations, details in SI). This explains how the hydrogenation process and surface plasmon resonance are crucial for the reaction at room temperature and pressure.

## Conclusion

We investigated $AuRu_{0.1}$, $AuRu_{0.2}$, and $AuRu_{0.3}$ bimetallic alloys for photocatalytic ammonia synthesis at ambient temperature and pressure under visible light illumination. The highest reaction rates and external quantum efficiency are observed on $AuRu_{0.2}$, which is ~ 60 μmol/g/h and ~0.12%. Such reactivity originates from the balancing of plasmonic and catalytic effects of the respective elements. The observation of an optimal alloy concentration along with semi-classical modeling suggests that surface-assisted Landau damping, rather than interband transitions, is the major contributor to the reaction. Using IR camera observations, we see that the light-driven reaction achieves higher reactivity than the thermal-driven reactions while operating at lower temperatures. *In-situ* DRIFTS measurements reveal that $NH_3$ decomposition at higher temperatures may be the limiting factor in the thermal reactions, whereas light-driven reactions exhibit increased $NH_x^*$ and $NH_3^*$ concentration across the laser powers studied. Combined with quantum mechanical simulations, our results suggest that plasmon-mediated hot carriers can accelerate the ammonia synthesis process by tailoring the hydrogenation and desorption of intermediate nitrogen species at room temperature and pressure. Therefore, light alone or $H_2$ alone cannot achieve $N_2$ dissociation, but the combination of light and $H_2$ can accelerate $N_2$ dissociation – the key bottleneck to breaking the triple bond of $N_2$. Interestingly, this mechanism is consistent with recent hypothesis of how nitrogenase enzymes catalyze ammonia production (using MoFe proteins) at mild conditions.[40] Our work provides guidance for more efficient plasmonic catalyst design for sustainable ammonia synthesis, potentially by tunning the hot electron densities from the surface effect, controlled Ru active centers[41,42] with increased surface area, and the light-driven enhancement of hydrogenation.

## Methods
**Preparation of AuRu bimetallic alloy and catalyst bed.**

The FCC phase of $AuRu_{0.1}$, $AuRu_{0.2}$, and $AuRu_{0.3}$ nanocrystals were prepared using a polyol reduction method with modification.[18,19] First, the precursors for Au and Ru in forms of hydrogen tetrabromoaurate (III) hydrate ($HAuBr_4 \cdot nH_2O$, 99.99%, Thermo Fisher Scientific, 55.56 mg) and potassium pentachloronitrosylruthenate (II) ($K_2Ru(NO)Cl_5$, Sigma-Aldrich, 40*x mg, where x represents the molar ratio of Ru), were dissolved in 10 mL diethylene glycol (Thermo Fisher Scientific) and stirred in the dark environment overnight. Following this, PVP (444 mg, MW=40,000) was dissolved in 100 mL ethylene glycol (Sigma-Aldrich), and the solution was kept stirring and heated up to 190 °C. The precursor solution was injected at 1.5 mL/min speed into the hot EG solution with a syringe infusion pump. After the precursor was fully injected, the solution was kept at the same temperature for another 15 min. The color transition to dark red or slight purple can be observed during the injection process. After the reaction stopped and cooled down to room temperature, the nanocrystals were separated by mixing the solvent with acetone and diethyl ether at the ratio of EG:acetone:diethyl ether=1:1.25:1 and ~8000 rpm first. Then the nanocrystal solution was redispersed into DI water and washed 3 times at ~135,000 rpm to get rid of the impurities of solvents and extra ligands.

The catalyst bed was prepared by the wet impregnation method. The 3 wt% AuRu nanocrystals were dropped into MgO aqueous solution and stirred overnight to mix fully with the support, and then the supported catalyst bed was centrifuged to collect the powder and dry the solvent. The powder was put into a furnace and heated at 350°C in air to remove the surfactants. The final powder was ground in a pestle and stored. All the catalytic measurements are reduced in a 5 sccm $N_2$ and 15 sccm $H_2$ environment at 350 °C for 1 hour to reduce the possible surface oxide.

**Characterization**

The X-ray diffraction (XRD) was implemented with an Empyrean X-ray Diffractometer from PANalytical with a Cu source (8.04 keV). The sample was prepared by drop-casting the colloidal solution on a clean sodium glass. ICP-OES measurements were conducted by a Thermo Scientific ICAP 6300 Duo View Spectrometer. TEM images and selected area electron diffraction were taken from an FEI Titan environmental TEM operated at 300 kV and recorded by a Gatan OneView camera. HADDF-STEM images and STEM-EDS elementary mappings were taken from a Thermo Fisher Spectra TEM at 300 kV.

**Catalytic measurements**

The catalyst bed is composed of 3 wt% of $AuRu_{0.1}$, $AuRu_{0.2}$, $AuRu_{0.3}$, and Au nanocrystals supported on MgO. In each experiment, 15 mg of catalyst bed was loaded into a customized stainless-steel Harrick high-temperature reaction chamber with Silco Tek coatings. 5 sccm $N_2$ and 15 sccm $H_2$ were constantly flowed into the reaction chamber. All the catalyst beds were heated up to 350 °C and kept for 1 hr to remove the potential surface oxide. The products in the outlet of the gas line were flowed into an acid trap composed of 0.1 M HCl. Optical illumination was performed with an NKT superK white laser equipped with a tunable Varia bandpass filter (bandwidth was set as ±25 nm for each selected wavelength). Surface temperatures of the catalyst beds were monitored using a thermography infrared camera (FLIR A700) with a temporal resolution of 30 frames per second and a spatial resolution of ~ 100 μm.

The catalytic rate is quantified by Nessler's reagent method, and evidenced by indophenol blue and $^1$H NMR measurements using the previously published protocol.[18,29]

(i) *Nessler method*

10 mL $NH_4Cl$ solution at the serial concentration of 0.22, 0.44, 0.55, 0.88, 1.1, 2.2, 3.3, 4.4, 5.5 μg/mL were mixed with 0.5 mL potassium sodium tartrate solution (0.5 g/mL) and 0.5 mL Nessler reagent. The mixed solutions were in a dark environment for 30 min. 2 mL of each solution was taken out, and measured the UV-Vis absorbance spectra from 400-500 nm. The absorbance at 420 nm was chosen to correlate the concentration according to the literature (Fig. S6). The products at the 10 mL acid trap were mixed with 0.5 mL potassium sodium tartrate and 0.5 mL Nessler reagent for the same courtesy to quantify the catalytic reaction rate.

(ii) *Indophenol blue method*

10 mL $NH_4Cl$ at the serial concentrations of 0.22, 0.44, 0.55, 0.88, 1.1, 2.2, 3.3, 4.4, 5.5 μg/mL were mixed with 2 mL of 1 M NaOH solution containing 5% salicylic acid and 5% sodium citrate. Subsequently, 1 mL of 0.05 M NaClO and 0.2 mL of 1% $C_5FeN_6Na_2O·2H_2O$ were added to the above solution. After mixing and storing in a dark environment for 1 h, UV-Vis absorbance spectra from 500-850 nm were taken and the peak at 680 nm was used to correlate the concentration of ammonia. The products at the 10 mL acid trap were tested using the same courtesy as the Nessler method as the calibration to quantify the catalytic reaction rate (Fig. S6 and S7).

(iii) *$^1$H NMR method*

$^1$H NMR spectra of the photocatalytic product, standard $NH_4Cl$, and $^{15}NH_4Cl$ were taken by a Brucker Neo 500 MHz NMR spectrometer (Fig. S7). 0.9 mL solution in the acid trap after the reaction was mixed with 0.1 mL DMSO-$d_6$ for the measurement.

**In-situ Diffuse Reflective Infrared Fourier Transform Spectroscopy (DRIFTS)**

*In-situ* DRIFTS spectra were carried out using the same gas flow, heating/light irradiation, as the photocatalysis measurements. A praying-mantis DRIFTS accessory is equipped with the Harrick cell on a Thermo Fisher Nicolet IS-50 spectrometer. The praying mantis accessory was purged with clean, dry air overnight before the measurement. A HdCdTe (MCT) detector operated at $LN_2$ temperature was employed to collect the spectra. Before in-situ reactions, a background spectrum was collected by averaging 16 scans. All spectra were presented in the form of absorbance according to -log(I/$I_0$), where $I_0$ and I are the spectra of background and during reactions, respectively.

**FDTD simulations**

The Extinction spectra (Fig. 1h) and the electric field for hot carrier cross-section calculations (Fig. 2f, 2g, and S9) were simulated using the FDTD method (Ansys Lumerical FDTD solutions). The single spherical particle model at a diameter ranging from 5 to 30 nm was chosen according to the size distribution from TEM images (Fig. S1 and S2). The Extinction spectrum of each particle was calculated by the broad-band multi-frequency TFSF approach with normal incident light, and the absorption/scattering cross-section was calculated by power flux of total field through a closed surface. The electric field enhancements and the refractive index on the surface of AuRu bimetallic alloy were recorded in the simulation. The hot carrier density was calculated based on the formula (1) to (4) and was implemented by a custom-written script for each model. The final spectra are the weighted average calculation results considering the size distribution from TEM images imported into a custom-written MATLAB code.

The dielectric permittivity of AuRu bimetallic alloy was treated using effective medium approximation (EMA),[43-45] where the Au dielectric permittivity was taken from Johnson and Christy,[46] and Ru dielectric permittivity was taken from a previously reported literature also fit it with Drude-Lorentz model.[47]

**Quantum Mechanical Calculations**

   **Periodic Plane-wave Density-Functional Theory (PW-DFT).** We carried out periodic PW-DFT calculations with the Perdew-Burke-Ernzerhof[48] (PBE) exchange-correlation (XC) functional. We used Grimme's D3 dispersion corrections with the Becke-Johnson (BJ) damping function[49,50] to correct the deficiencies in PBE for non-covalent interactions. We applied the projector augmented-wave (PAW) method[51] as implemented in the Vienna Ab Initio Simulation Package (VASP) version 5.4,[52] in which we explicitly treat the 1s electron for H, the 2s and 2p electrons for N, the 4d and 5s electrons for Ru, and the 5d and 6s electrons for Au. We used a plane-wave (PW) basis kinetic energy cutoff of 660 eV with a Methfessel-Paxton Fermi smearing width of 0.09 eV during the structure optimizations. We used a single Γ point to perform the electronic integrations. We used a four-layer Ru-doped Au(111) (RuAu$_{95}$ per computational unit cell) slab in the calculations to model the surface of the Ru-doped Au nanoparticle. Between the slabs, we added a 15-Å-thick vacuum layer. We applied a dipole correction along the z-axis direction to eliminate unphysical interactions between periodic images.[53] We optimized the minimum-energy paths (MEPs) by using the climbing image nudged elastic band (CI-NEB) method.[54] The maximum of the atomic forces on the images on the NEB is converged to within 0.03 eV/Å.

   **Embedded Correlated Wavefunction (ECW) Theory.** In the ECW theory,[55-58] the whole periodic system is partitioned into a cluster and its environment, and the interaction between the cluster and environment is given by the embedding potential $V_{emb}$, which is computed at the periodic PW-DFT level. Here, we used an 11-atom cluster (RuAu$_{10}$), which was carved out from the Ru-doped Au slab, with the PBE XC functional and the PW basis to calculate the $V_{emb}$. The final ECW energy $E^{ECW}$ is calculated as $E^{ECW} = E_{tot}^{PW\text{-}DFT} + \left(E_{clst}^{CW} - E_{clst}^{DFT}\right)$, where $E_{tot}^{PW\text{-}DFT}$ is the PW-PBE energy (without the D3-BJ correction) for the total periodic system, performed on the previously optimized geometries along the MEPs. The energy correction term, i.e., $\left(E_{clst}^{CW} - E_{clst}^{DFT}\right)$, is computed by using the cluster model with the atom-centered Gaussian-type basis set in the presence of the embedding potential. In the CW calculation or DFT calculation, the original electronic Hamiltonian $H_0$ becomes $H_0 + V_{emb}$ to include the interactions between the bare cluster and its environment. The adsorbate was then added onto the embedded bare metal cluster with its geometry (internal coordinates) fixed at the corresponding PW-PBE-D3BJ optimized structures along the MEPs. $E_{clst}^{DFT}$ was computed with the PBE XC functional (without the D3-BJ correction) with $V_{emb}$. The basis sets used for the cluster calculations were the ANO-RCC-VTZP basis.[59-61] For $E_{clst}^{CW}$, we performed embedded complete active space second-order perturbation theory (CASPT2).[62-64] The ground-state wavefunctions were obtained by using the single-state embedded complete active space self-consistent field (emb-CASSCF) method.[65] In particular, for surface-bound $H_2$ dissociation, we used a (10e, 10o) active space, which includes $H_2$ $\sigma$ and $\sigma^*$ orbitals, and 8 frontier orbitals (4 occupied and 4 empty) on the metallic cluster that are of s/d hybridized character. For $N_2$ direct dissociation, we used a (14e, 14o) active space, which includes $N_2$ ($\sigma_{2p}, \sigma_{2p}^*$), two pairs of ($\pi, \pi^*$), and 8 frontier orbitals (4 occupied and 4 empty) on the metallic cluster. For H-facilitated $N_2$ dissociation, we partitioned a large active space of (22e, 22o) using the two-particle two-hole embedded restricted active space self-consistent field (emb-RASSCF) method:[66] (14e, 7o) in RAS1 (including two H 1s orbitals, Ru $d$ – N $p$ $\sigma$ and $\pi$ orbitals, and frontier orbitals on the metallic surface), 7 correlating orbitals in RAS3, and (8e, 8o) in RAS2, which consists of $N_2$ ($\sigma_{2p}, \sigma_{2p}^*$), two pairs of ($\pi, \pi^*$), and the ($\pi_{Rud-Np}, \pi'_{Rud-Np}$) pair.

The dynamical correlations were recovered by using ground-state emb-CASPT2 theory calculations, which were performed with the single-state emb-CASSCF wavefunctions, with an IPEA shift[64] of 0.25 hartree. We included all of the excitations from all of the electrons (excluding the inner core electrons of Au (from 1s to 4s) and Ru (from 1s to 3s)) and all of the orbitals (active, inactive, and virtual orbitals) at the second-order perturbation level. We optimized the excited-state wavefunctions using the embedded state-averaged complete active space self-consistent field (emb-SA-CASSCF) method;[67] the SA calculations performed here always included the ground state. To obtain a balanced treatment for all of the states, in the 10-state emb-SA-CASSCF calculations, we used an equal weight for each root in the optimization. The excitation energies were computed by emb-CASPT2 with the wavefunctions optimized by emb-SA-CASSCF. An imaginary shift of 0.2 hartree was applied to avoid intruder states.[68] We added the excitation energies onto the ground-state single-state embCASPT2 potential energy to obtain the excited-state PESs. We used our in-house modified OpenMolcas program[69] to perform all the ECW calculations on the cluster.

## Acknowledgments

All authors at Stanford acknowledge the support from Keck Foundation under grant No.994816, the Office of Basic Energy Sciences, US Department of Energy, Division of Materials Science and Engineering, DE-AC02-76SF00515. We also acknowledge the support from ATW-Alan T Waterman Award from National Science Foundation under grant No. 1933624, and the support from the NSF, Center for Adopting Flaws as Features (NSF CHE-2124983). Y.W. and A. M. acknowledges the support from the Office of Naval Research MURI Award N00014-21-1-2377. J.L.B. acknowledges the financial support provided by the American Chemical Society Petroleum Research Fund (PRF no. 65744-DNI6). In addition, J.L.B. thanks the Boston College Linux Cluster Center for cluster computing resources. B.B.B. was supported by the National Science Foundation

Graduate Research Fellowship under grant number DGE-1656518. We acknowledge the use and support of the Stanford Nano Shared Facilities (SNSF), supported by the National Science Foundation under award ECCS-2026822. L.Y. acknowledges the helpful discussion regarding synthesis and collection of AuRu bimetallic alloy with Dr. Quan Zhang from Kyoto University, Japan.

**Author Contributions**

L. Y. and J. A. D. conceptualized the research, including the design of electromagnetic simulations and experiments. L. Y., B. B. B., A. X. D., and Z. C. finalized the synthetic protocol and experimental details under the supervision of M. R. J. and J. A. D. L. Y., B. B. B., A. X. D., A. S. M.-G., Y. C., K. X., and Y. W. conducted all TEM imaging and analysis, supervised by Y. C., A. M., and J. A. D. L. Y. performed the electromagnetic simulations and calculations. E. B. and J. L. B. carried out the first-principles QM calculations, ECW excited-state calculations, and provided insights into reaction pathways. L. Y. and Y. Z. conducted the in-situ DRIFTS measurements and analysis, while L. Y. and Z. X. performed the synchrotron X-ray absorption measurements. All authors contributed to discussions, provided insights, and participated in manuscript preparation.

# Supporting Information

**External quantum efficiency (EQE) for ammonia production at different wavelengths**

The EQE was acquired by performing the ammonia synthesis under a super continuum laser equipped with a bandpass filter with a 50 nm bandwidth centered at different wavelengths (490, 510, 525, 550, 575, 600, 625, 650, 675, and 700 nm, respectively). The power was fixed and measured by an optical power meter. The formula used to calculate of EQE is:

$$EQE = \frac{Number\ of\ electrons\ responsible\ for\ ammonia\ generation}{Number\ of\ input\ photons} \times 100\% = \frac{3 \times N_A \times n_{NH_3}}{\frac{Input\ power\ from\ laser \times time\ of\ illumination}{\hbar\omega}} \times 100\%$$

In this equation, $N_A$ is Avogadro constant, $n_{NH_3}$ is the mole quantity of produced ammonia, the input power from laser is fixed at 100 mW, and $\hbar\omega$ is the energy quantum of the photon with the different center wavelengths (corresponding to the frequency $\omega$).

**Supporting Figures and Tables**

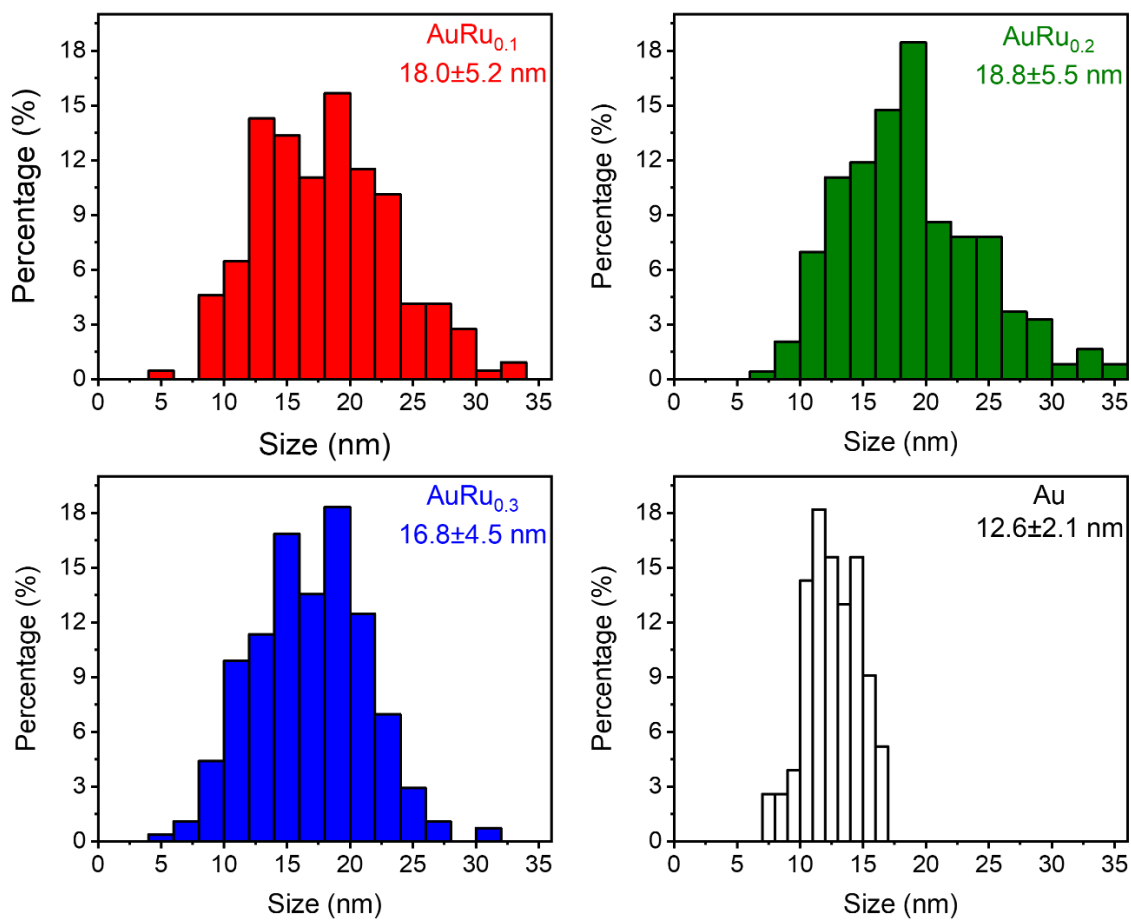

**Fig. S1| Size distribution histograms for AuRu$_{0.1}$, AuRu$_{0.2}$, AuRu$_{0.3}$, and Au nanocrystals.** Each size distribution counts 5 random spots in TEM samples with over 250 nanoparticles.

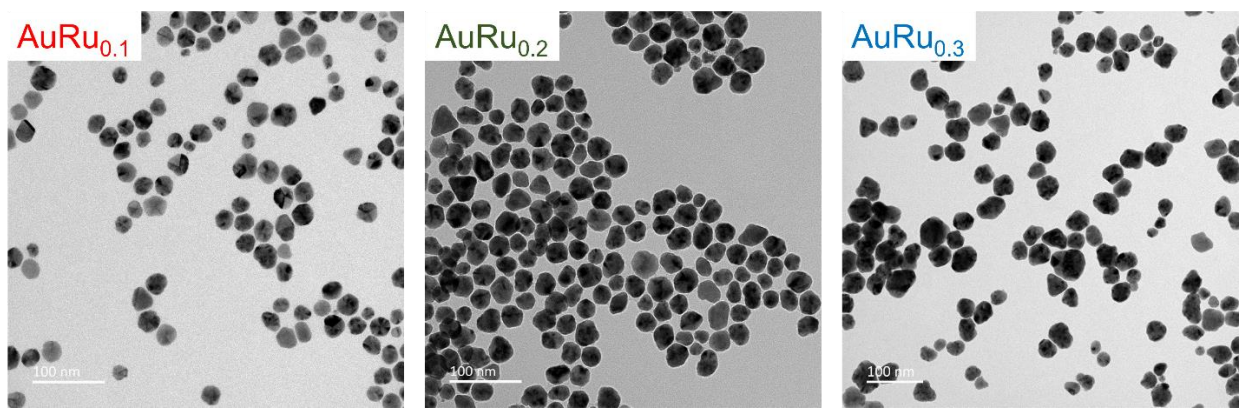

**Fig. S2| Representative TEM images of AuRu0.1, AuRu0.2, and AuRu0.3.** Scale bars: 100 nm

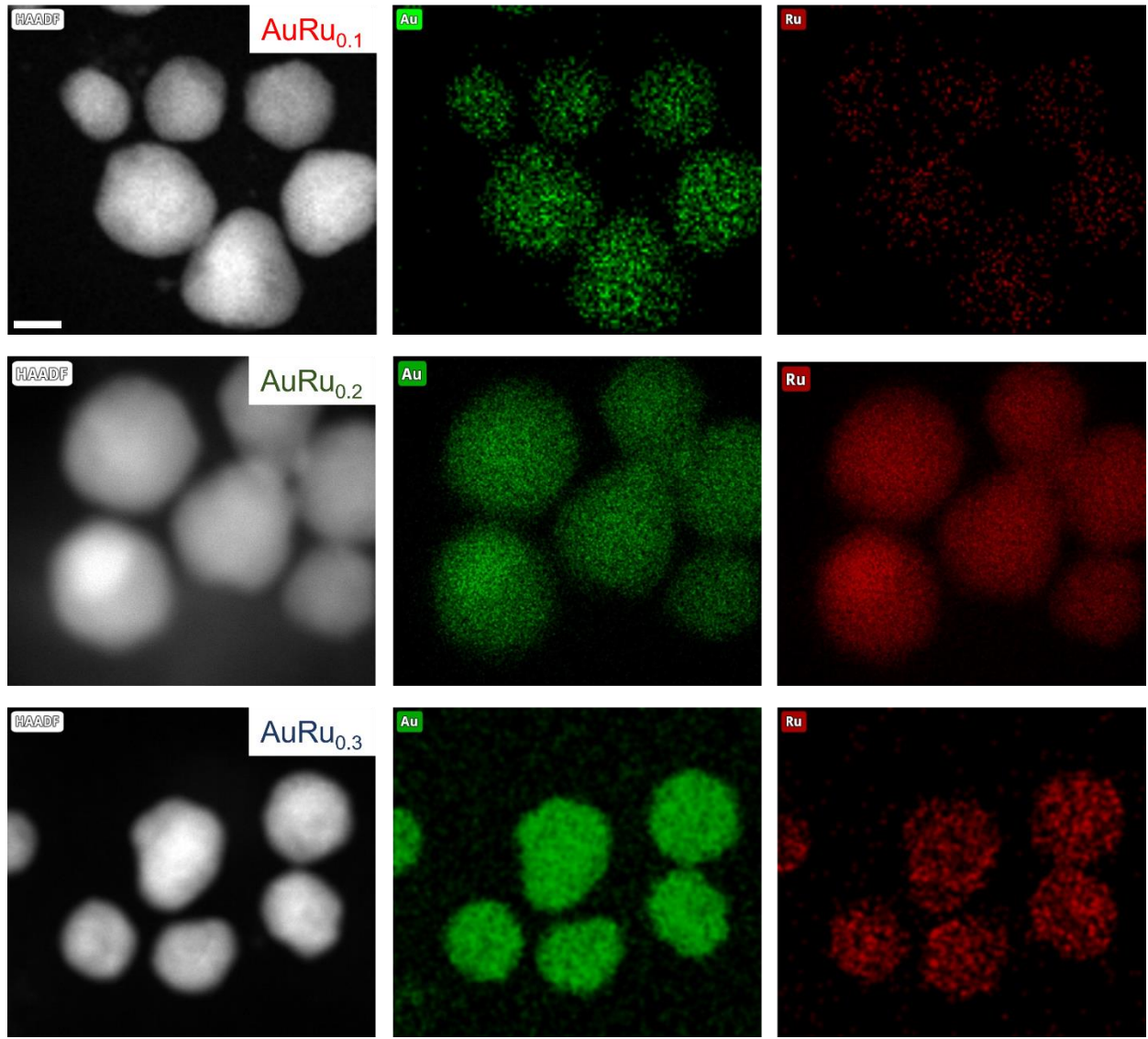

**Fig. S3| Elementary mapping of Au and Ru from energy dispersive X-ray spectroscopy of scanning transmission electron microscopy (STEM-EDX) of AuRu$_{0.1}$, AuRu$_{0.2}$, and AuRu$_{0.3}$ nanocrystals.** Scale bars for all images: 10 nm

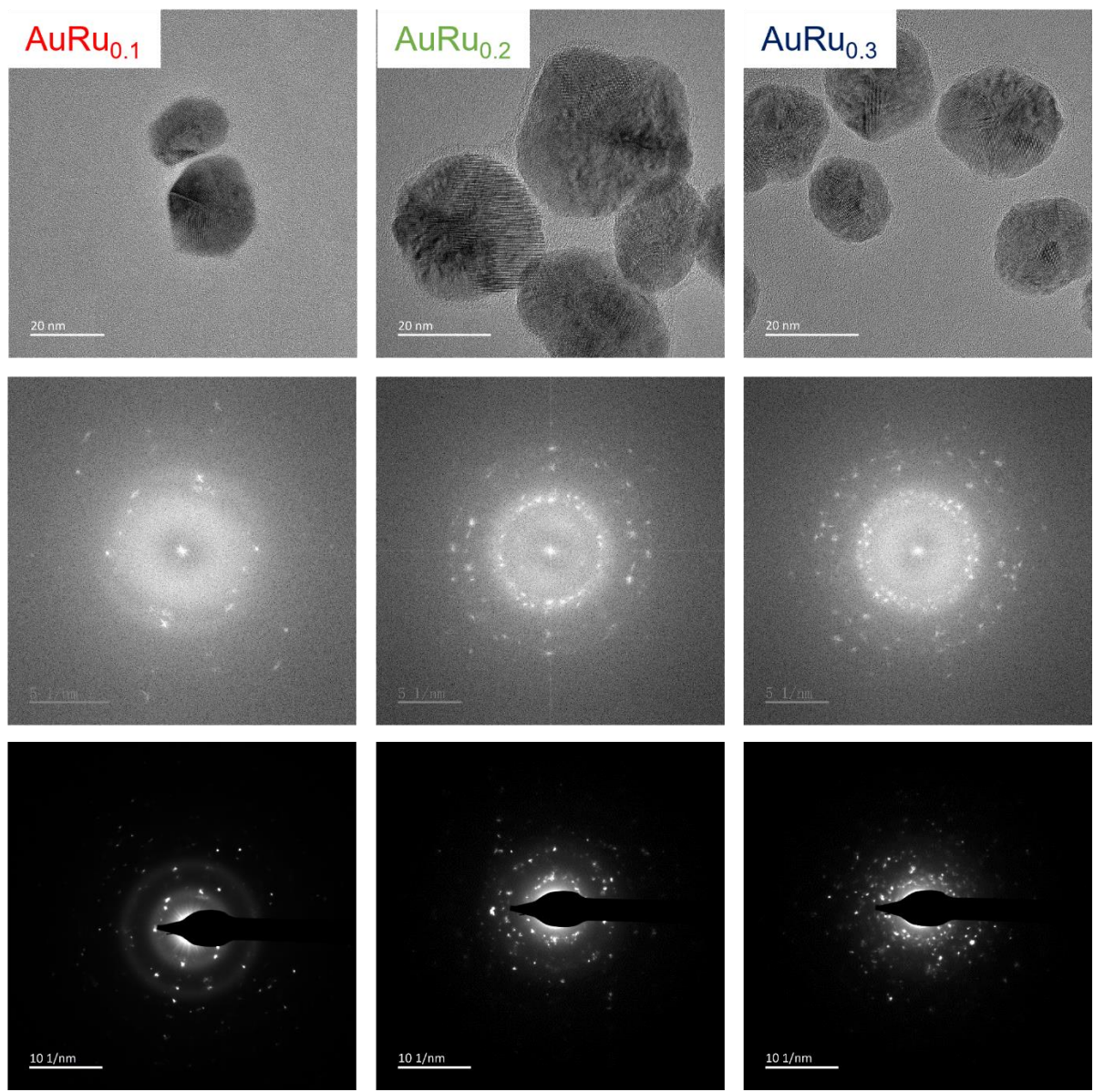

**Fig. S4| High-Resolution transmission electron microscopy images (HRTEM), its corresponding Fast Fourier Transform (FFT) and selected area electron diffraction (SAED) pattern of AuRu$_{0.1}$, AuRu$_{0.2}$, and AuRu$_{0.3}$ nanocrystals.**

| Samples | Au mass concentration | Ru mass concentration |
|---|---|---|
| AuRu$_{0.1}$ | 0.9501 | 0.0391 |
| AuRu$_{0.2}$ | 1.269 | 0.1269 |
| AuRu$_{0.3}$ | 1.025 | 0.1901 |

**Table S1| ICP-OES results of AuRu atomic ratio.**

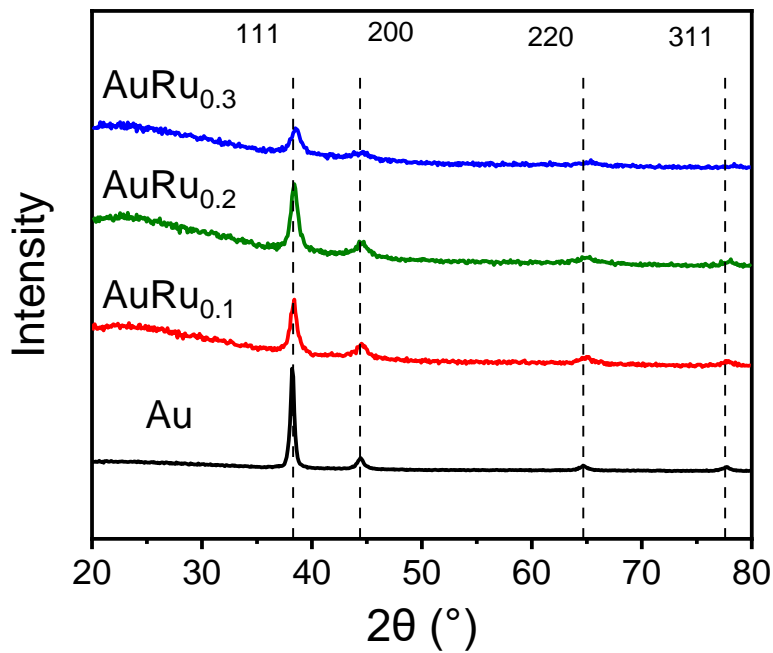

**Fig. S5| X-ray diffraction spectra (XRD) of Au, AuRu$_{0.1}$, AuRu$_{0.2}$, and AuRu$_{0.3}$ nanocrystals.**

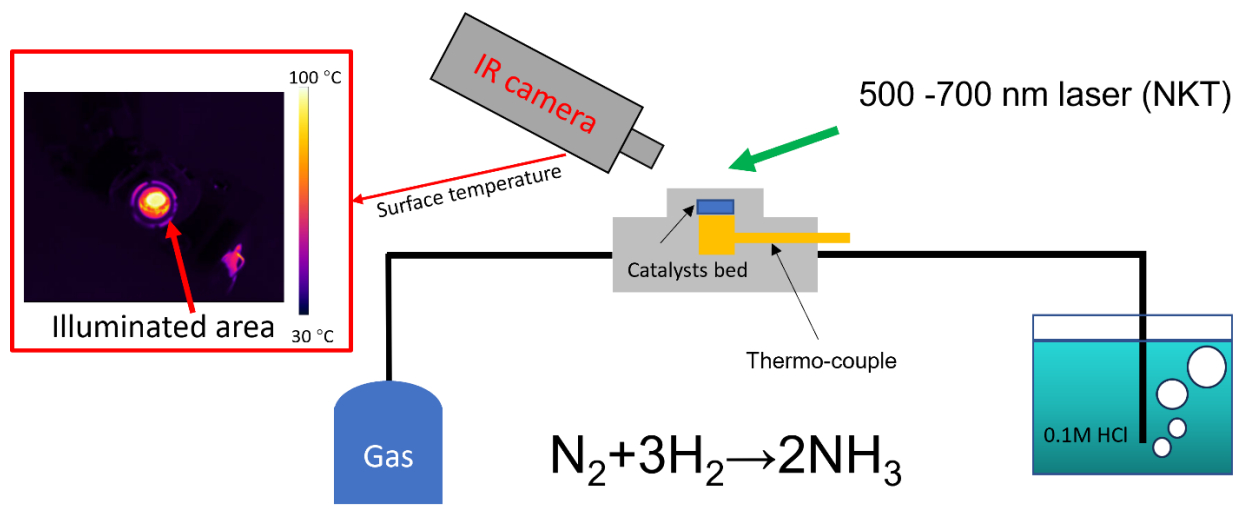

**Fig. S6| Schematic of the setup for ammonia synthesis.** The N$_2$ and H$_2$ gas were constantly flowed through the Harrick reaction chamber with the ratio of 1:3 and the total flow rate at 20 sccm.

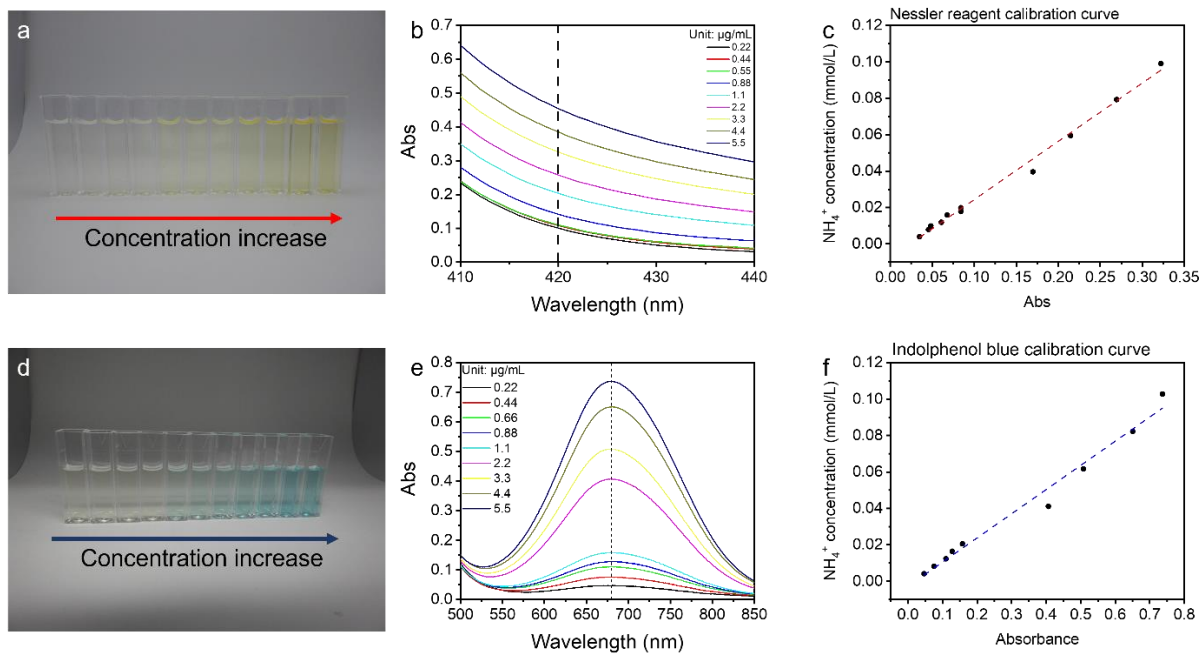

**Fig. S7| Calibration of Nessler reagent and Indophenol blue.** (a) A photograph of a series concentration of standard $NH_4Cl$ with Nessler method. (b) UV-Vis absorbance spectra of the serial standard $NH_4Cl$. (c) Calibration curve of the ammonia concentration with the absorbance of Nessler reagent at 420 nm. (d) a photograph of a series concentration of standard $NH_4Cl$ with Indophenol blue method. (e) UV-Vis absorbance spectra of the serial standard $NH_4Cl$. (f) Calibration curve of the ammonia concentration with the absorbance of Indophenol blue at 680 nm.

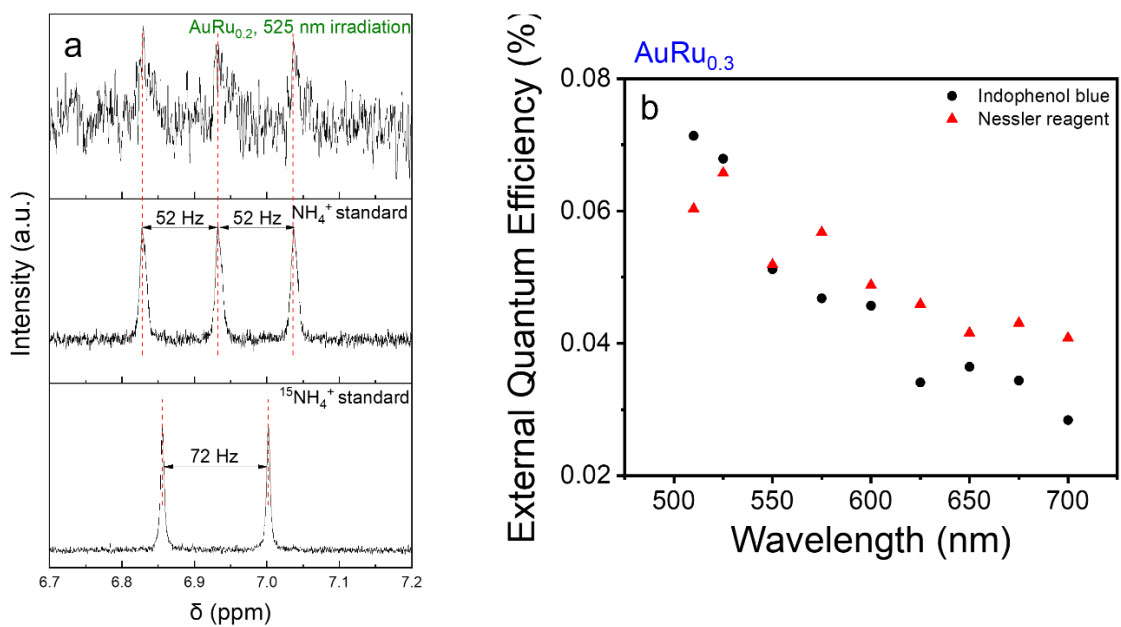

**Fig. S8| Detection of ammonia product in acid trap.** (a) $^1$H NMR results of the product on AuRu$_{0.2}$, and its comparison with standard NH$_4$Cl, and $^{15}$NH$_4$Cl aqueous solution. (b) A representative wavelength-dependent measurements using both Nessler method and Indophenol blue method for the same batch of AuRu$_{0.3}$ catalysts.

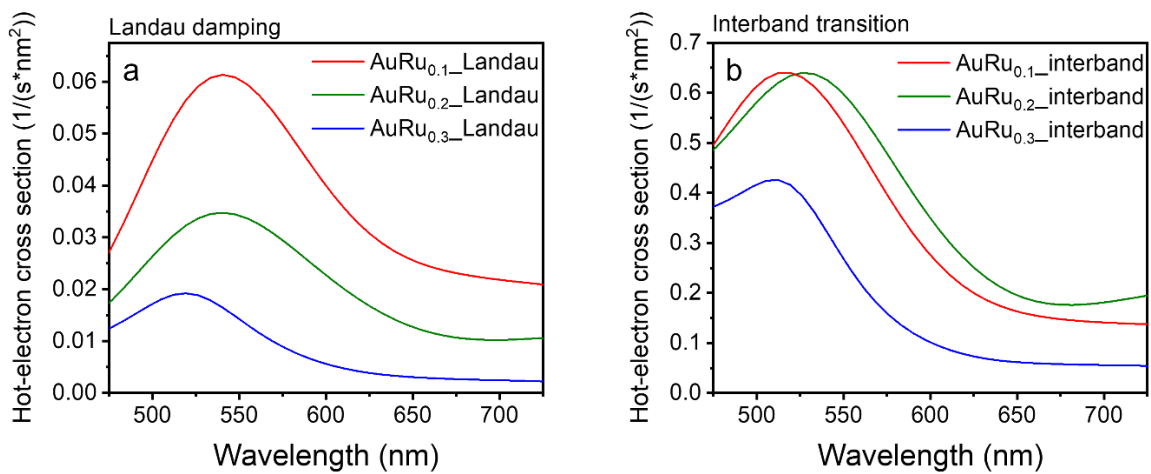

**Fig. S9| Theoretical calculations of hot-carrier cross section from (a) Landau damping and (b) interband transition.**

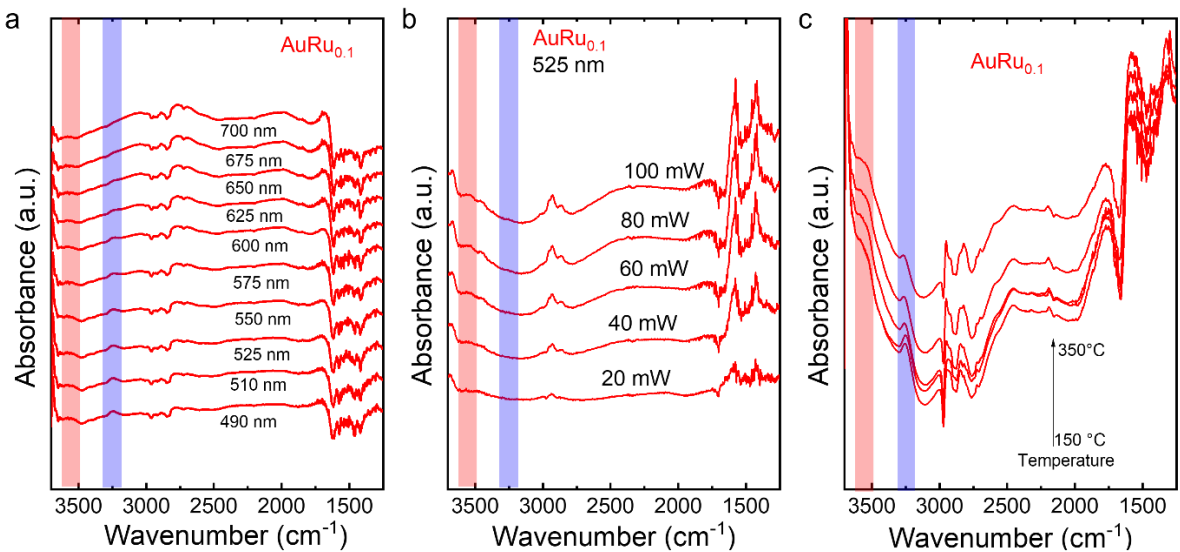

**Fig. S10|** *In-situ* DRIFTS spectra of (a) Wavelength-dependent, (b) power-dependent, and (c) thermal-driven reaction on AuRu$_{0.1}$.

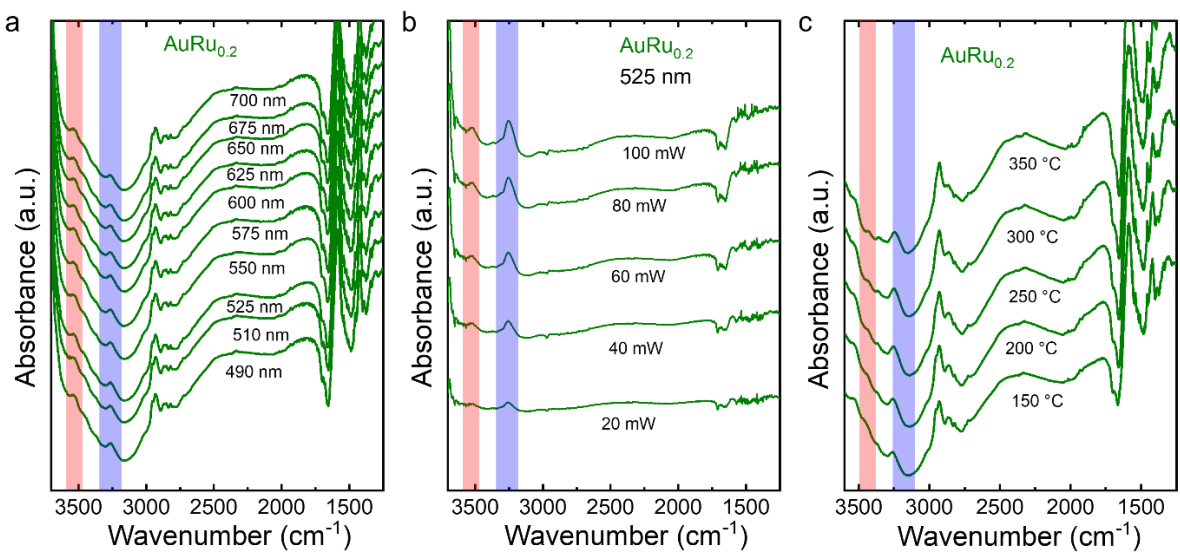

**Fig. S11|** *In-situ* DRIFTS spectra of (a) Wavelength-dependent, (b) power-dependent, and (c) thermal-driven reaction on AuRu$_{0.2}$.

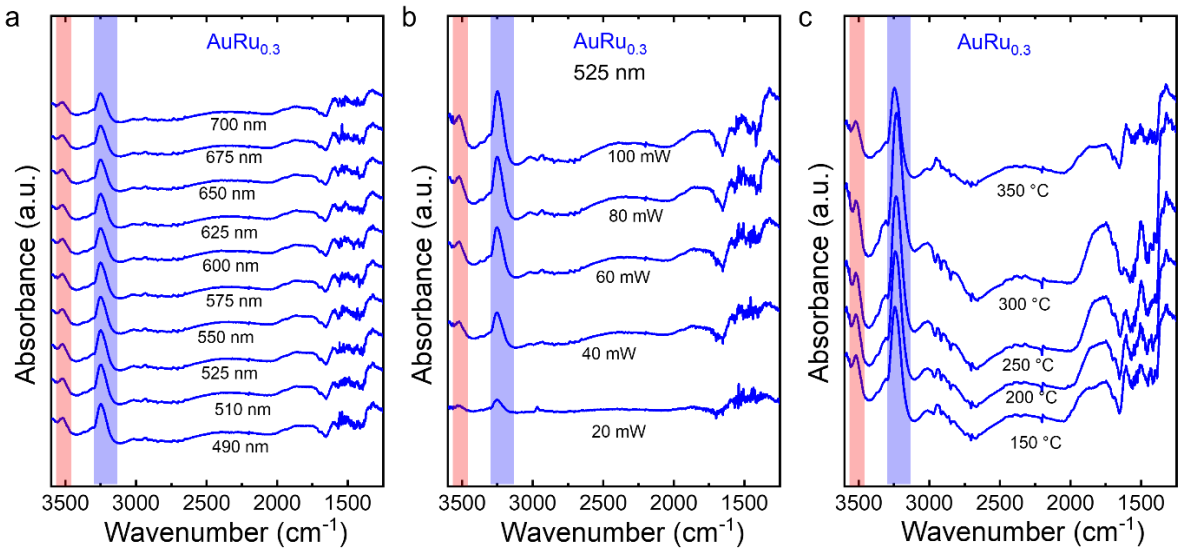

**Fig. S12|** *In-situ* DRIFTS spectra of **(a)** Wavelength-dependent, **(b)** power-dependent, and **(c)** thermal-driven reaction on AuRu$_{0.3}$.

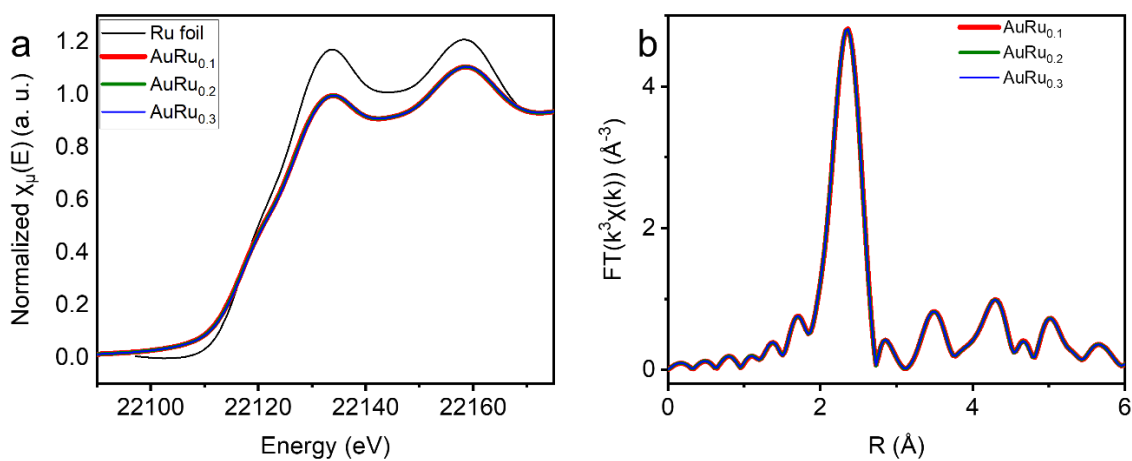

**Fig. S13| Synchrotron X-ray Absorption spectra. (a)** XANES spectra at the Ru K edge of Ru foil, AuRu$_{0.1}$, AuRu$_{0.2}$, and AuRu$_{0.3}$. **(b)** k$^3$-weighted EXAFS spectra of AuRu$_{0.1}$, AuRu$_{0.2}$, and AuRu$_{0.3}$.

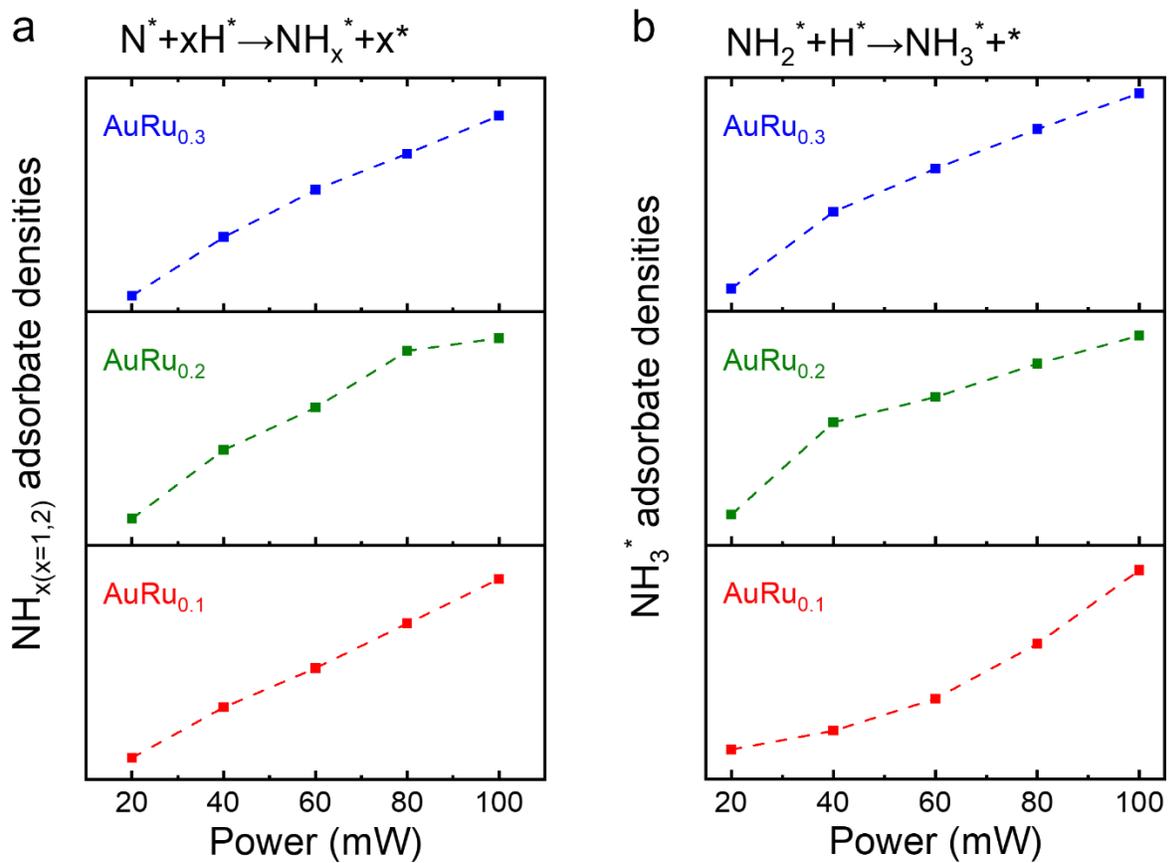

**Fig. S14| Power-dependent adsorbate densities of $NH_x^*$ and $NH_3^*$ species on $AuRu_{0.1}$, $AuRu_{0.2}$, and $AuRu_{0.3}$.**

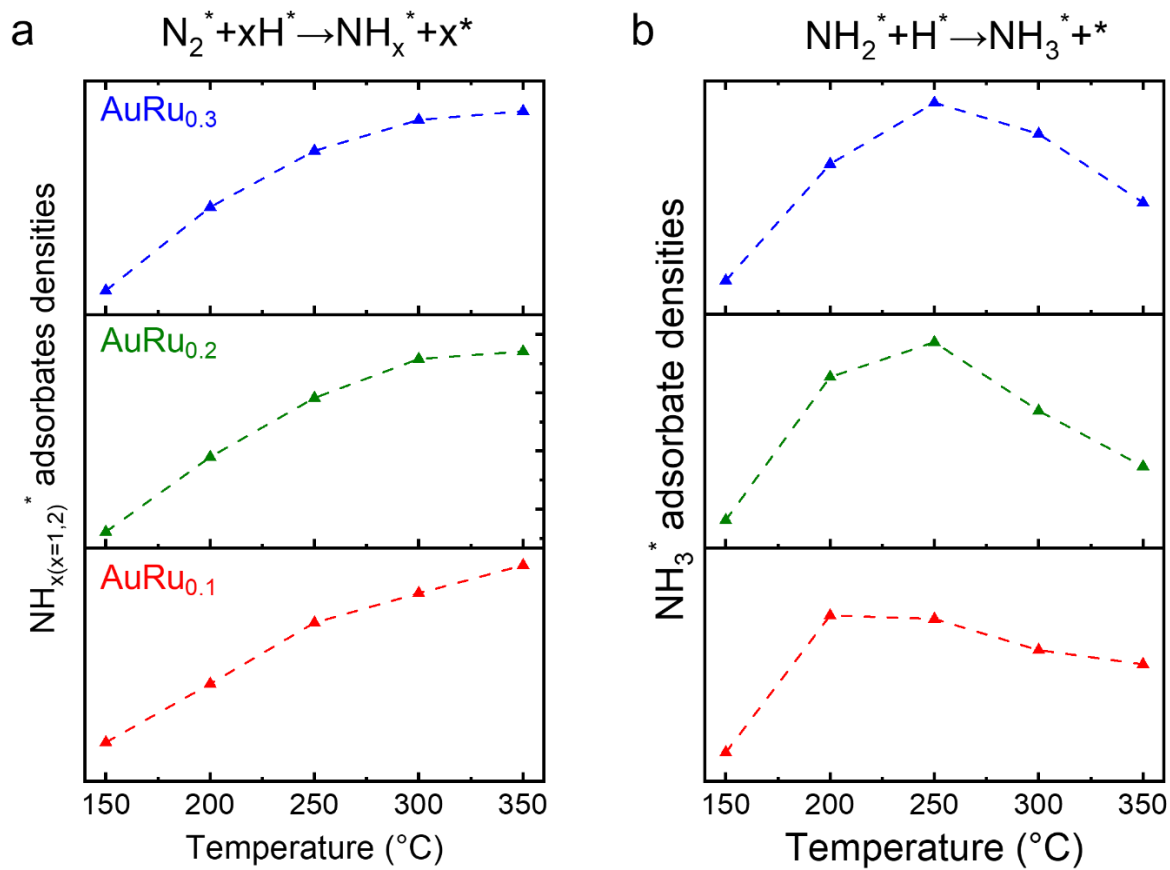

**Fig. S15| Thermal-driven adsorbate densities of $NH_x^*$ and $NH_3^*$ species on $AuRu_{0.1}$, $AuRu_{0.2}$, and $AuRu_{0.3}$.**

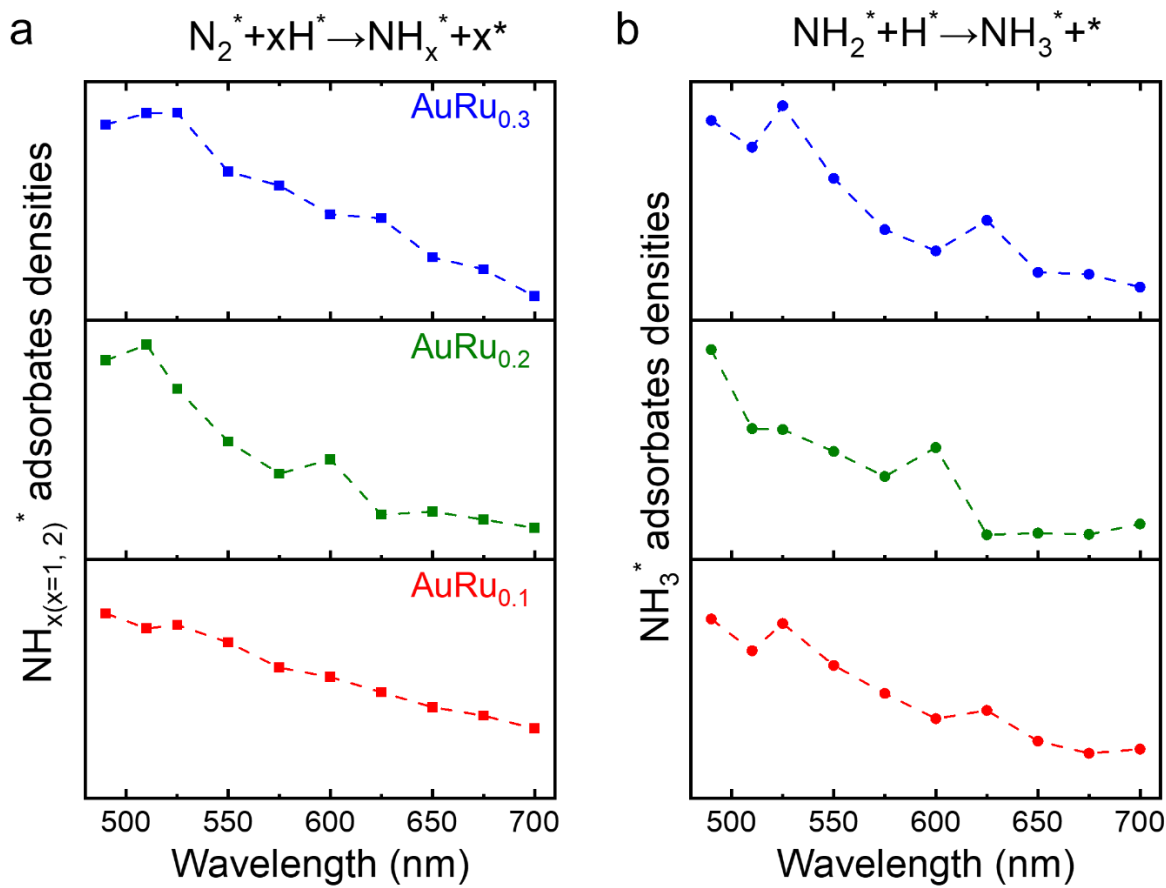

**Fig. S16| Wavelength-dependent adsorbate densities of $NH_x^*$ and $NH_3^*$ species on $AuRu_{0.1}$, $AuRu_{0.2}$, and $AuRu_{0.3}$.**

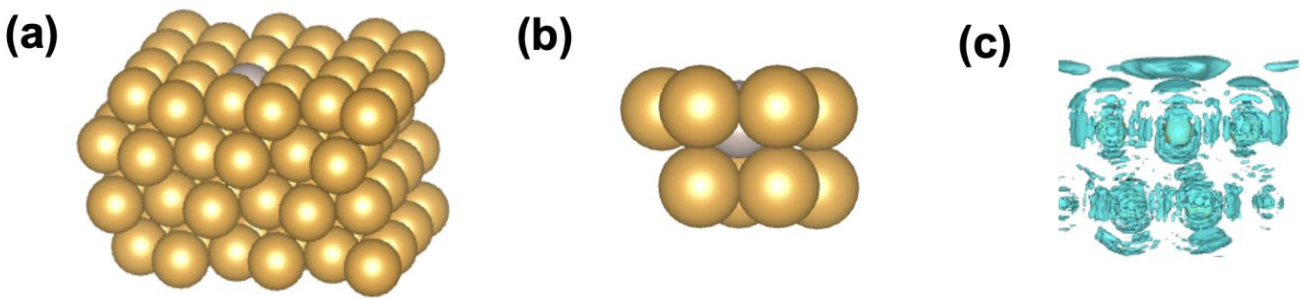

**Figure S17| Optimization of the geometry in the ECW model. (a)** Four-layer Ru-doped Au (111) periodic slab model (with $RuAu_{95}$ per computational unit cell), in which the yellow balls are Au atoms and the grey ball is the Ru atom. **(b)** The carved $RuCu_{10}$ cluster for the embedded cluster calculations. **(c)** The converged optimized embedding potential $V_{emb}(r)$.

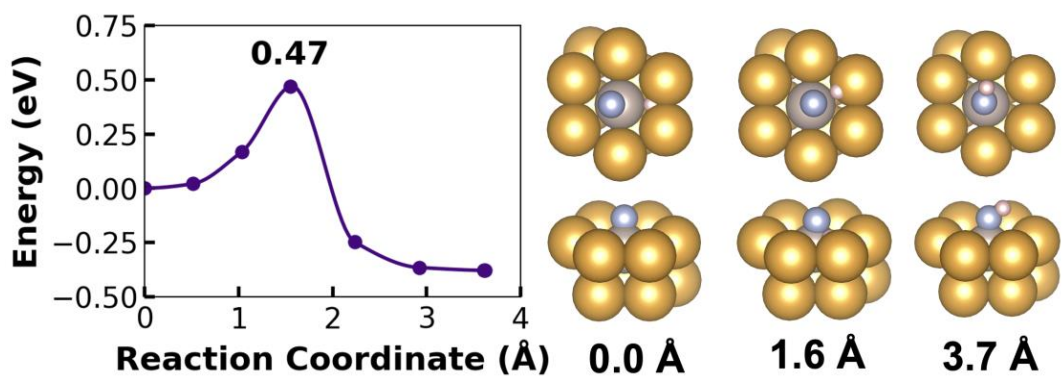

**Figure S18| Surface-bound N-H bond formation: N(ads) + H (ads) → NH(ads). (Left panel)** The relative energies (eV) computed at the plane-wave (PW) DFT PBE-D3BJ level for the ground-state minimum-energy path (MEP) as a function of the reaction coordinate. **(Right panel)** The top and side views of the optimized critical structures on the ground-state MEP with the corresponding reaction coordinates.

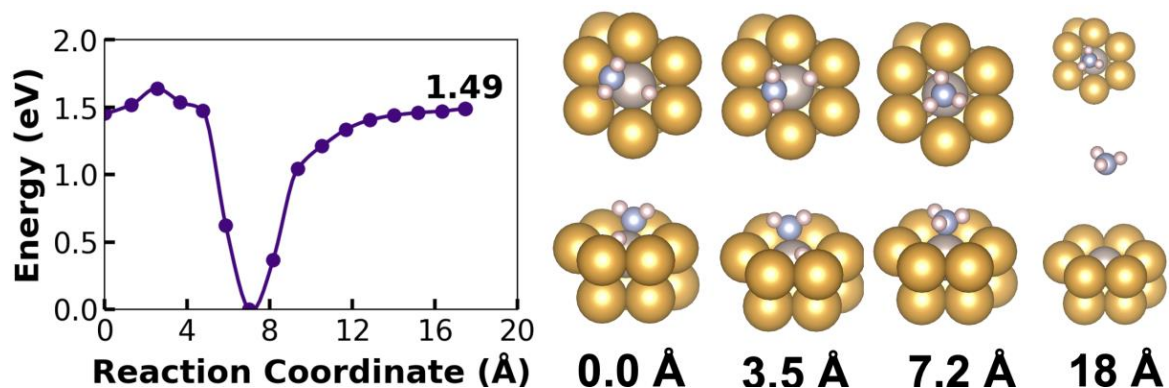

**Figure S19| $NH_3$ associative desorption: $NH_2$ (ads) + H (ads) → $NH_3$ (ads) → $NH_3$ (gas). (Left panel)** The relative energies (eV) computed at the plane-wave (PW) DFT PBE-D3BJ level for the ground-state minimum-energy path (MEP) as a function of the reaction coordinate. **(Right panel)** The top and side views of the optimized critical structures on the ground-state MEP with the corresponding reaction coordinates.

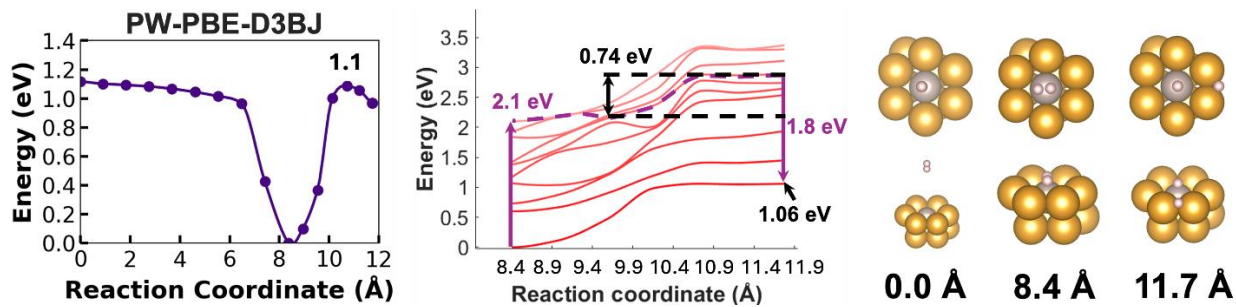

**Figure S20|.** The relative energies (eV) along the reaction coordinate (Å) for the $H_2$ dissociative adsorption pathway: $H_2$ (gas) → $H_2$ (ads) → 2H (ads). **(Left panel)** The relative energies (eV) computed at the plane-wave (PW) DFT PBE-D3BJ level for the ground-state minimum-energy path (MEP) as a function of the reaction coordinate. **(Middle panel)** The relative energies

computed at the ECW emb-CASPT2 level for surface-bound $H_2$ dissociation $H_2$ (ads) → 2H (ads), including both the ground state and excited states. The solid arrows represent the LSPR absorption and vertical excitation/de-excitation. The dashed lines represent the reactive pathways on the excited-state potential-energy surfaces. **(Right panel)** The top and side views of the optimized critical structures on the ground-state MEP with the corresponding reaction coordinates.

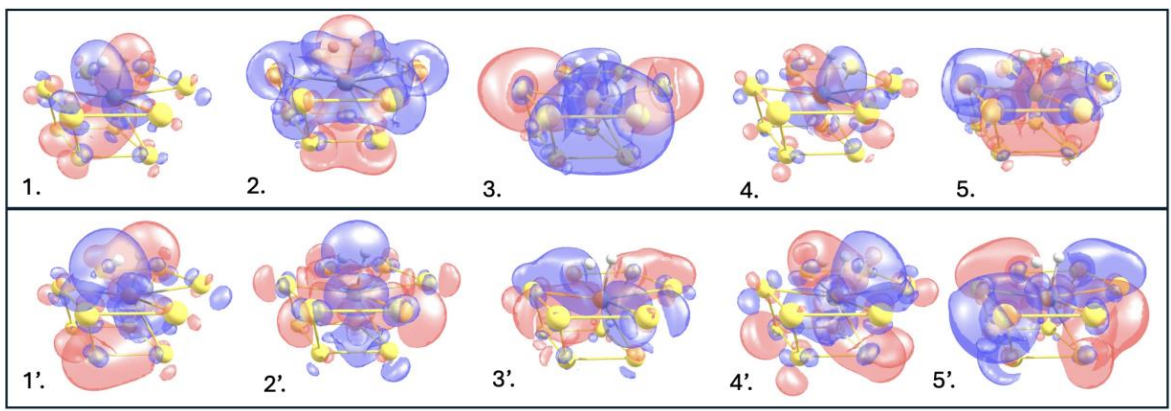

**Figure S21|** Optimized emb-CASSCF ground-state active orbitals for the adsorbed $H_2$ (ads) on the carved cluster $RuAu_{10}$. The active space includes 10 electrons and 10 orbitals. Orbital 2 is the $H_2$ $\sigma$ orbital, and orbital 2′ is the correlating $\sigma^*$ orbital. Orbitals 1, 3–5 are primarily the surface cluster orbitals of s/d hybridized characters, and orbitals 1′, 3′–5′ are the corresponding correlating orbitals.

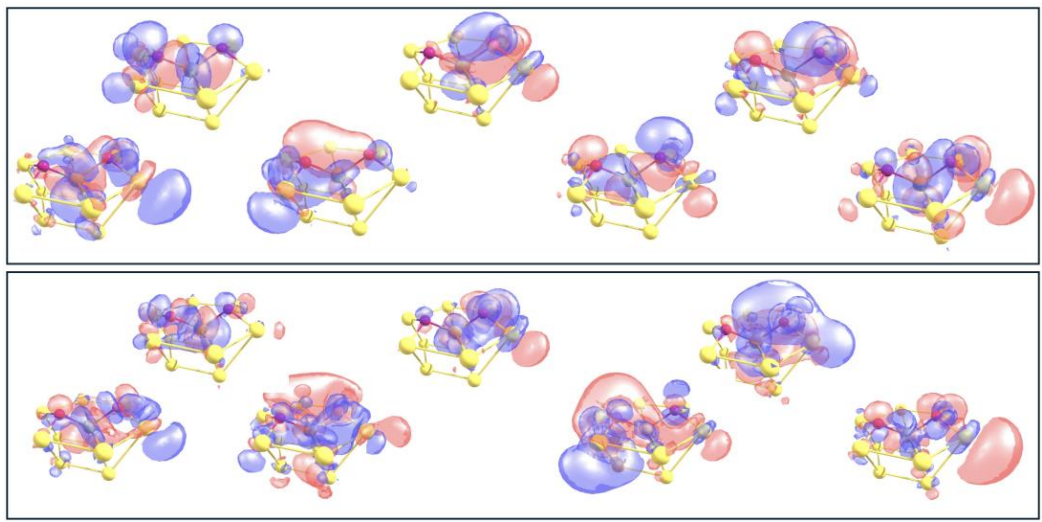

**Figure S22|** Optimized emb-CASSCF ground-state active orbitals for the surface-bound dissociated 2N (ads) on the carved cluster $RuAu_{10}$ from the $N_2$ direct dissociation pathway. The active space includes 14 electrons and 14 orbitals.

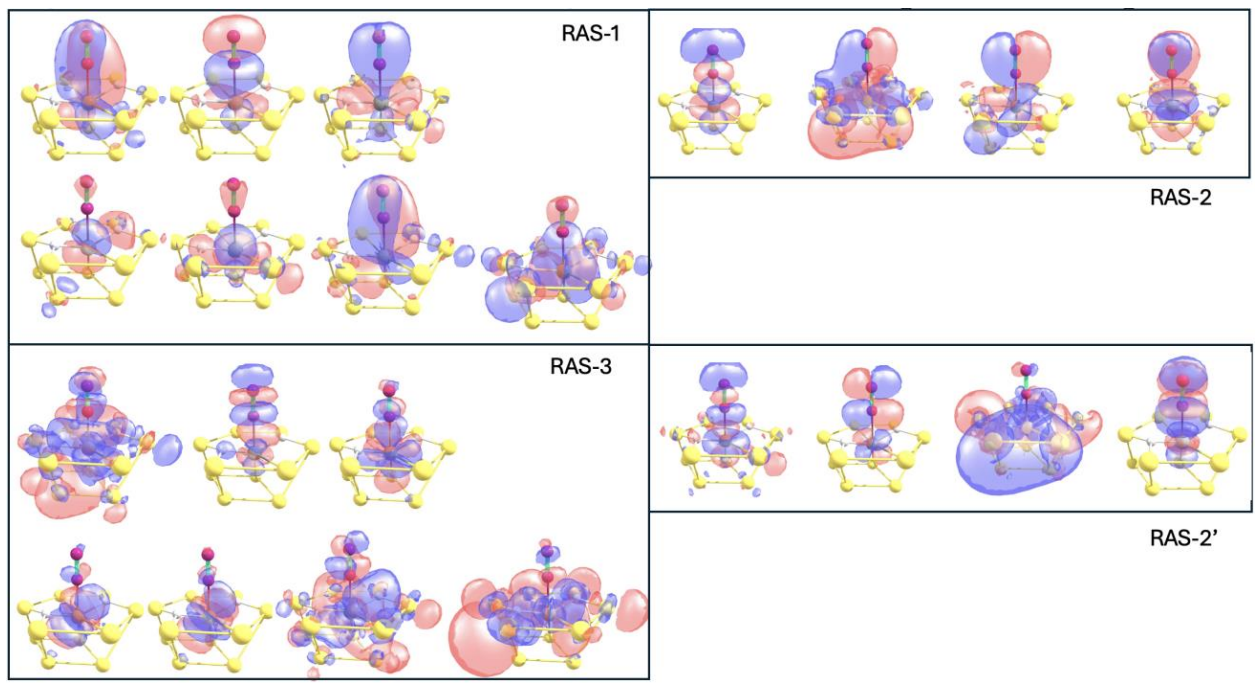

**Figure S23|** Optimized emb-RASSCF ground-state active orbitals for the surface-bound H (ads) + N$_2$ (ads, vertical) configuration on the carved cluster RuAu$_{10}$, which is the initial state of the H-facilitated N$_2$ dissociation pathway: 2H (ads) + N$_2$ (ads, vertical) → N$_2$H$_2$ (ads) → 2NH (ads). The RAS1 space includes 14 electrons in 7 active orbitals, and the RAS3 space includes 7 active orbitals. The RAS2 space includes 8 electrons and 8 orbitals consisting of N$_2$ ($\sigma_{2p}$, $\sigma_{2p}^*$), two pairs of ($\pi$, $\pi^*$), and the ($\pi_{\text{Ru}d-\text{N}p}$, $\pi'_{\text{Ru}d-\text{N}p}$) pair. We allow two-particle two-hole transitions.

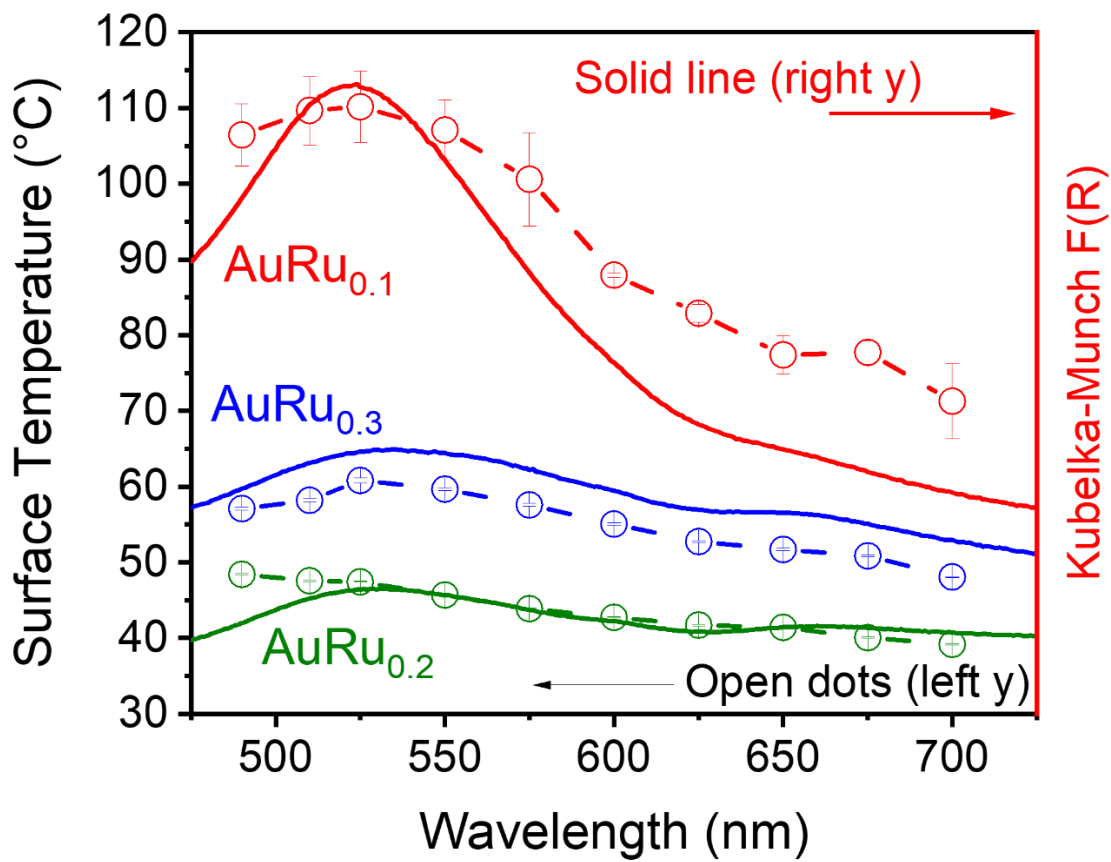

**Fig. S24|** Wavelength-dependent surface temperature captured by a real-time infrared camera on AuRu$_{0.1}$, AuRu$_{0.2}$, and AuRu$_{0.3}$.

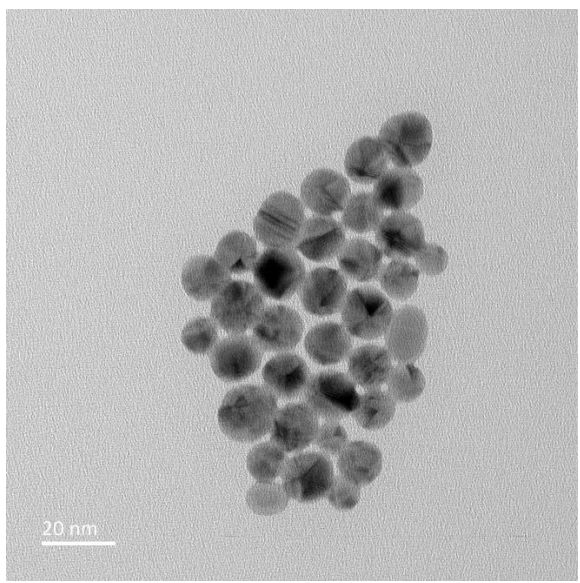
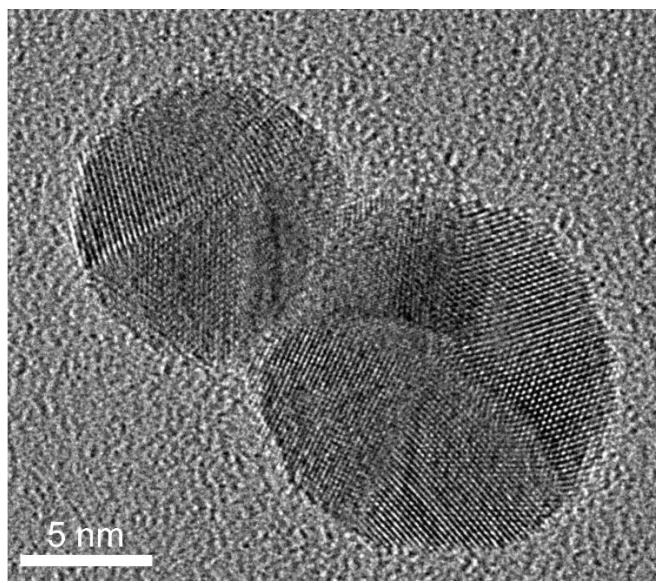

**Fig. S25| High-resolution TEM images of Au nanocrystals as the control group.**

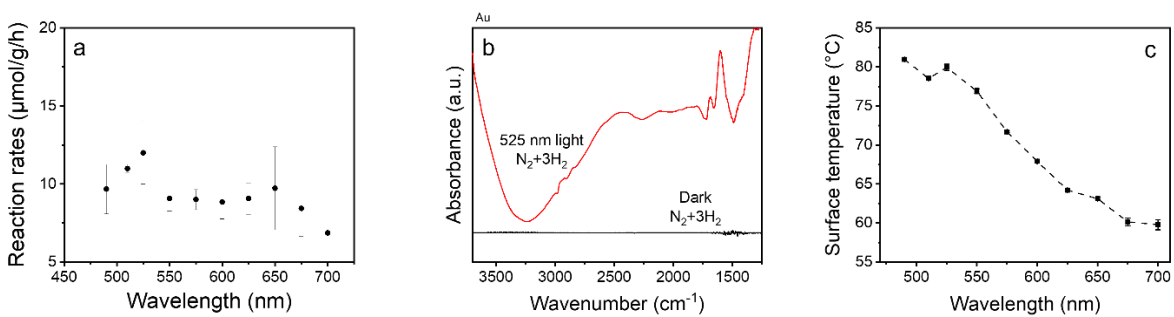

**Fig. S26| (a) Wavelength-dependent reactivity of ammonia synthesis, (b) in-situ DRIFTS spectra, and (c) wavelength-dependent surface temperature captured by infrared camera on 3 wt% Au supported on MgO.**

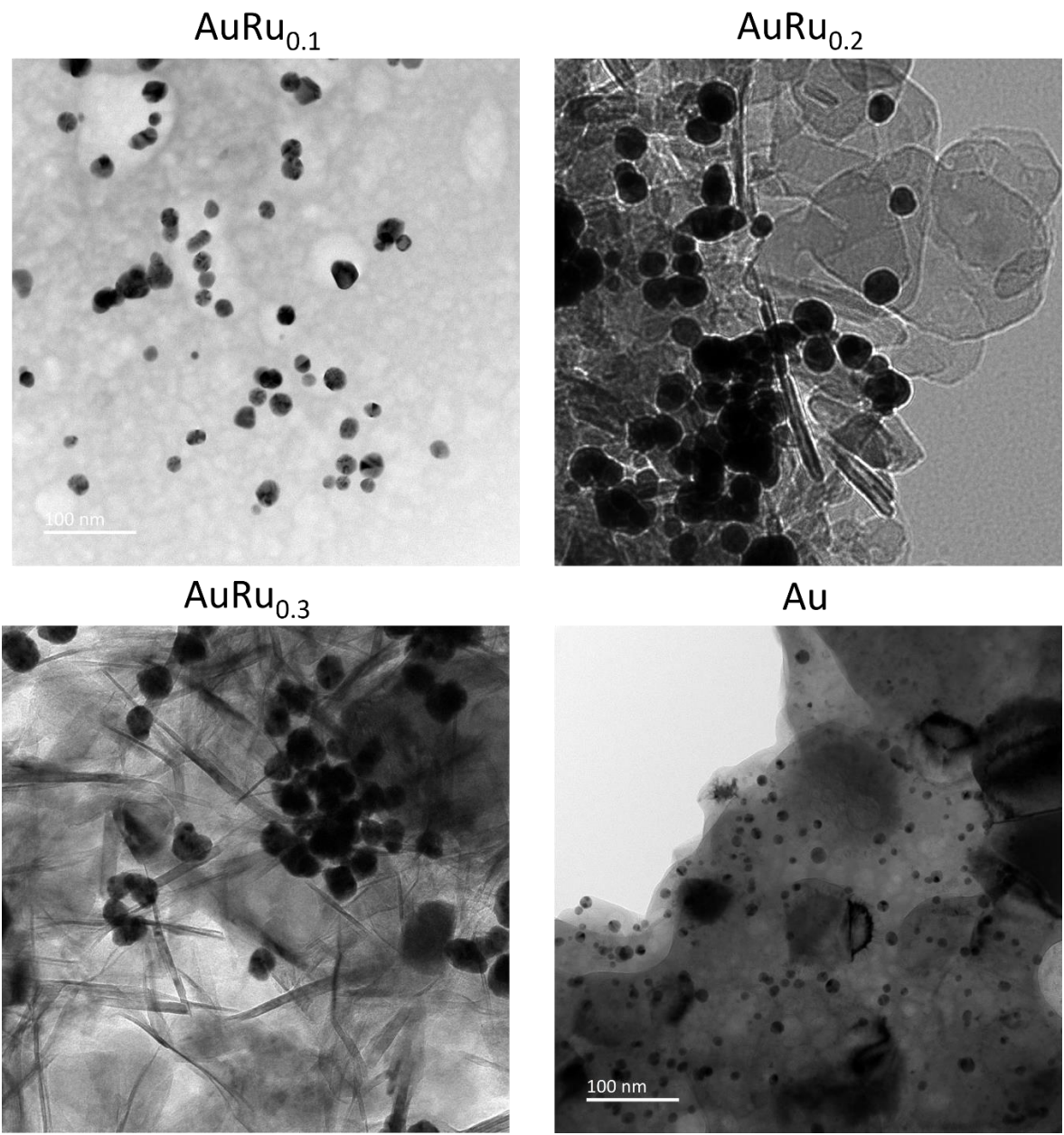

**Fig. S27| TEM images of Au, AuRu$_{0.1}$, AuRu$_{0.2}$, and AuRu$_{0.3}$ nanocrystals supported on MgO.**

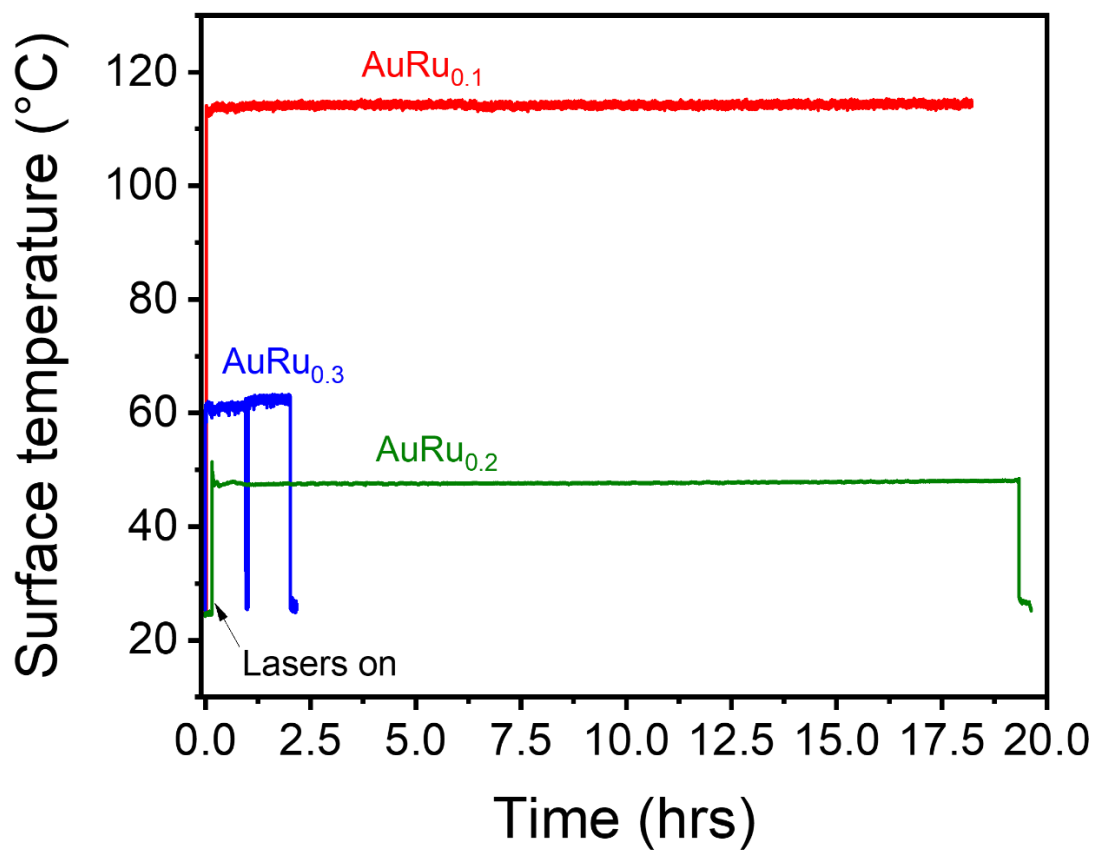

**Fig. S28| Surface temperature of AuRu$_{0.1}$, AuRu$_{0.2}$, and AuRu$_{0.3}$ at 525 nm and 100 mW/cm$^2$ for long-time running under 5 sccm N$_2$ and 15 sccm H$_2$ environment.**

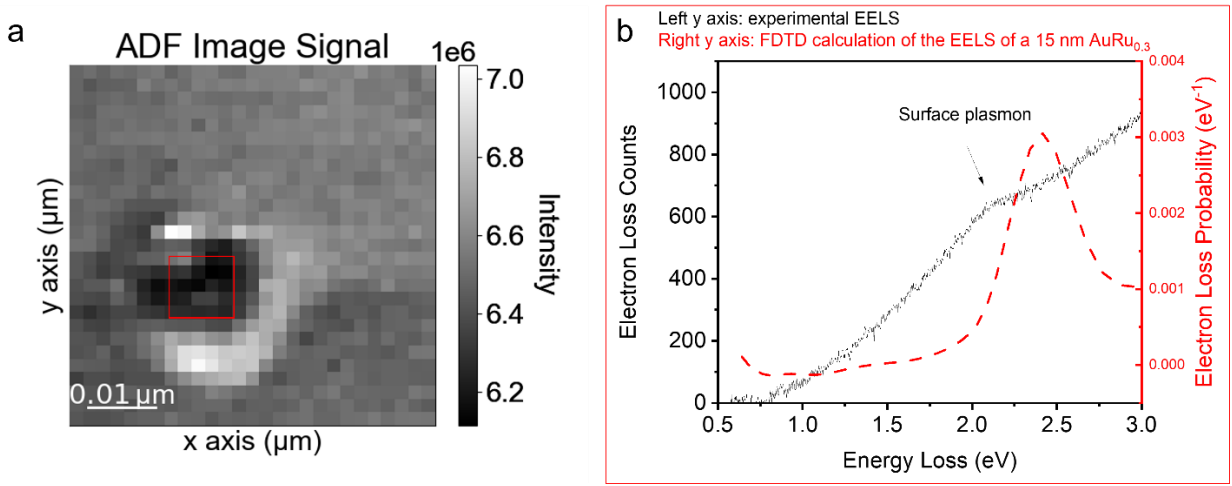

**Fig. S29| Electron Energy Loss Spectroscopy (EELS) of a representative AuRu$_{0.3}$ nanocrystal.** (a) Annular Dark Field-Scanning Transmission Electron Microscopic (ADF-STEM) image at 60 kV of the AuRu$_{0.3}$ nanocrystal. The rectangle represents the area of the integration of EELS signal in (b). (b) The EEL spectroscopy at visible range (Left y axis), and its corresponding simulated Electron Loss probability by a FDTD method.

**The Summary of Computational Results**
**(1) Thermal catalysis**

| Elementary step | DFT Barrier | ECW Barrier | Comment |
|---|---|---|---|
| $H_2$ dissociative adsorption: $H_2$ (gas) → $H_2$ (ads) → 2H (ads) | 1.1 eV | 1.1 eV | |
| $N_2$ dissociative adsorption: $N_2$ (gas) → $N_2$ (ads, vertical) → $N_2$ (ads, horizontal) → 2N (ads) | 1.1 eV<br>5.0 eV | 6.2 eV | This is NOT the dominant $N_2$ dissociation mechanism. |
| H-facilitated $N_2$ dissociation: 2H (ads) + $N_2$ (gas) → 2H (ads) + $N_2$ (ads, vertical) → $N_2H_2$ (ads) → 2NH (ads) (→ N (ads) + H (ads) optional) | 2.0 eV<br>3.0 eV<br>(0.85 eV) | 2.2 eV<br>3.3 eV | This is the dominant $N_2$ dissociation mechanism. |
| $NH_3$ associative desorption: (Optional: N (ads) + H (ads) → NH (ads)) $NH_2$ (ads) + H (ads) → $NH_3$ (ads) → $NH_3$ (gas) | (0.47 eV)<br>0.2 eV<br>1.5 eV ($NH_3$ desorption energy) | | |

Note: DFT (Density functional theory) provides a low-accuracy estimation, while the ECW (embedded correlated wavefunction) theory provides a quantitative high-accuracy estimation of the energetics. We perform the high-accuracy ECW calculations on only a selected (rate-determining) number of DFT-optimized reaction pathways.

**(2) Photocatalysis** (with LSPR peak at ~ 2.0 – 2.2 eV)
Based on the results in thermal catalysis, we focus on the following rate-determining steps and explore the effective excited-state barriers at the high-accuracy ECW level.

| Elementary step | ECW Barrier (Ground-state chemistry) | ECW Barrier (Excited-state chemistry) | Comment |
|---|---|---|---|
| Surface-bound $H_2$ dissociation: $H_2$ (ads) → 2H (ads) | 1.1 eV | ~ 0.7 eV | |
| Surface-bound N2 dissociation: $N_2$ (ads, horizontal) → 2N (ads) | 6.2 eV | ~ 4.3 eV | This is NOT the dominant $N_2$ dissociation mechanism. |
| Surface-bound H-facilitated $N_2$ dissociation: 2H (ads) + $N_2$ (ads, vertical) → $N_2H_2$ (ads) → 2NH (ads) | 2.2 eV<br>3.3 eV | ~ 1.0 eV<br>~ 1.6 eV | This is the dominant $N_2$ dissociation mechanism. |